\documentclass[pra,twocolumn,preprintnumbers,amsmath,amssymb,superscriptaddress,longbilbiography,showpacs]{revtex4-1}
\usepackage{graphicx}
\usepackage{latexsym}
\usepackage{amsmath}
\usepackage{graphics}
\usepackage{amssymb}
\usepackage{subfigure}
\usepackage{layout}
\usepackage{verbatim}
\usepackage{amsfonts,epsfig,dsfont}
\usepackage{natbib}
\usepackage{hyperref}

\newcommand{\ket}[1]{\left|#1\right>}
\newcommand{\bra}[1]{\left< #1 \right|}
\newcommand{\beq}{\begin{equation}}
\newcommand{\eeq}{\end{equation}}
\newcommand{\bea}{\begin{eqnarray}}
\newcommand{\eea}{\end{eqnarray}}

\usepackage{color}

\usepackage{ulem} 

\begin{document}

\title{How to detect qubit-environment entanglement generated during qubit dephasing
}

\author{Katarzyna Roszak}
\affiliation{Department of Theoretical Physics, Faculty of Fundamental Problems of Technology, Wroc{\l}aw University of Science and Technology,
50-370 Wroc{\l}aw, Poland}
\affiliation{Institute of Physics, Academy of Sciences of the Czech Republic, 18221 Prague, Czech Republic}

\author{Damian Kwiatkowski}
\affiliation{Institute of Physics, Polish Academy of Sciences, Aleja Lotnikow 32/46, PL-02668 Warsaw, Poland}

\author{{\L}ukasz Cywi{\'n}ski}
\affiliation{Institute of Physics, Polish Academy of Sciences, Aleja Lotnikow 32/46, PL-02668 Warsaw, Poland}

\date{\today}

\begin{abstract}
	We propose a straightforward experimental protocol to test
whether qubit-environment entanglement is generated during pure dephasing of a qubit. The protocol is implemented using only measurements and operations on the qubit --
	 it does not involve the measurement of the system-environment state of interest, but the preparation and measurement of the qubit in two simple variations. 
	A difference in the time dependencies of qubit coherence between the two cases testifies to the presence of entanglement in the state of interest.
	Furthermore, it signifies that the environment-induced noise experienced by the qubit cannot be modeled as a classical stochastic process independent of the qubit state. 
	We demonstrate the operation of this protocol on a realistically modeled nitrogen vacancy center spin qubit in diamond interacting with a nuclear spin environment, and show that the generation of entanglement should be easily observable in this case.
\end{abstract}
\maketitle

\section{Introduction}
The interaction between a quantum system and its environment leads to decoherence \cite{Zurek_RMP03,Hornberger_LNP09} of superpositions of a system's pointer states \cite{Zurek_PRD82}. This ubiquitous feature of quantum open system dynamics has fundamental significance for the realistic description of all possible devices employing truly quantum features of physical systems for various tasks, as well as for the understanding of the quantum-classical transition \cite{Zurek_RMP03,Schlosshauer_book,Zurek_revisited}. The sensitivity of experimentally investigated qubits to environmental influence has also led to the development of a whole field of research devoted to the use of 
qubits to characterize their environments \cite{Degen_RMP17,Szankowski_JPCM17}. 

While any environment of a qubit should be in principle described quantum mechanically, it is now clear that environments relevant for the description of {\it pure dephasing} of qubits (for examples showing that pure dephasing is a very common dominant source of decoherence see e.g.~\cite{Nakamura_PRL02,Roszak_PRA06,Biercuk_Nature09,Bylander_NP11,Medford_PRL12,Staudacher_Science13,Muhonen_NN14,Malinowski_PRL17,Szankowski_JPCM17})
can often be modeled as sources of noise, the properties of which are {\it independent} of dynamics of these qubits, and even of their existence, see \cite{Degen_RMP17,Szankowski_JPCM17}.
We stress that this feature -- the ability to correctly describe the dynamics of 
the environment that leads to dephasing of the qubit by looking only at the dynamical properties of the environment, or in other words the absence of visible back-action of the qubit on the environment -- is taken here as the defining one of the ``classical environmental noise'' model of dephasing. 
Environments that have their dynamics unaffected by presence of the qubit can be modeled classically, by specifying all the multi-point correlation functions that characterize the stochastic process \cite{Szankowski_JPCM17,Norris_PRL16}, but note that the converse is not necessarily true: it could be possible to describe qubit decoherence caused by an environment by a model of external classical noise, while the actual joint qubit-environment evolution 
involves nontrivial back-action of the qubit on the environment and creation of quantum correlations between the two - in fact, pure dephasing of a freely evolving qubit (but, interestingly, not of a higher-dimensional system \cite{Helm_PRA09}) can always be effectively described by constructing an artificial model of external classical noise \cite{Landau_Streater_LAA93,Crow_PRA14}. 
Our goal here is not to show when one cannot come up with an {\it effectively classical} model of the environmental influence on the qubit, but to devise a simple experiment, the positive result of which clearly proves that treating the environment as independent of the qubit is impossible. The former is really a statement on the dynamics of the qubit (``can these dynamics of an open quantum system be reconstructed by introducing classical noise acting on the qubit''), while the latter is a statement on the physical nature of the environment coupled to the qubit. It is thus somewhat surprising that the experiment described here relies only on control and measurement of the qubit. These features make it of course easy to implement.

While the necessary conditions for applicability of such a ``classical noise'' approximation are not known, a large size of the environment and its high temperature are, as expected, positively correlated with ``classicality'' of qubit dephasing.
In the simplest - and very often realistic, as it arises for an environment consisting of many uncorrelated sub-environments, each weakly coupled to the qubit - case of noise with Gaussian statistics, full characterization of the noise is contained in its spectral density. In this case, qubits can be straightforwardly 
used as noise spectrometers \cite{Degen_RMP17,Szankowski_JPCM17}. 
It also should be noted that spectroscopy of non-Gaussian noise has been theoretically put forward in \cite{Norris_PRL16}, but the implementation of the protocol proposed there is definitely much more involved \cite{Sung_arXiv19} than in case of reconstruction of spectrum of Gaussian noise. 
However, while the classical noise model of qubit dephasing is believed to be widely applicable, a vexing fundamental problem remains unsolved: how can one unambiguously prove that decoherence of a given qubit is in fact {\it truly quantum}, i.e.~not amenable to description using classical environmental noise. 

The general issue of quantum vs.~classical nature of environmentally-induced qubit dephasing has another facet. For an initially pure state of the environment, the dephasing is in one-to-one correspondence with qubit-environment entanglement (QEE) generation \cite{Kuebler_AP73,Zurek_RMP03,Schlosshauer_book,Roszak_PRA15}. However, in the realistic case of a mixed initial state of the environment, decoherence {\it does not} have to be accompanied by generation of QEE \cite{Eisert_PRL02,Hilt_PRA09,Maziero_PRA10,Pernice_PRA11,Roszak_PRA15}. This fact is not strongly stressed in most of seminal papers on decoherence. The reasons were twofold. First, their focus was on the most strikingly quantum situation, in which coherence of one system is lost due to establishment of entanglement with a larger system that is also in pure state - the state of the whole remains pure, but its coherent nature becomes inaccessible by measurements on the system only. Second, the theory of mixed state entanglement started to be developed years after the foundations of decoherence theory (for historical perspective on them see \cite{Schlosshauer_book}) had been established. It is worth noting that the definition of separable mixed states (and the definition of entangled mixed states that follows from it) was given only in 1989 \cite{Werner_PRA89}.
After the theory of entanglement of mixed states was developed to a sufficiently advanced degree \cite{Plenio_QIC07,Horodecki_RMP09,Braunstein_RMP05,Aolita_RPP15}, the issue of system-environment entanglement generated during system's dephasing was revisited.
In Ref.~\cite{Eisert_PRL02} it was shown, that in the quantum Brownian motion model (an oscillator coupled to an environment of other oscillators), for a large class of initially mixed qubit states no system-environment entanglement was generated during decoherence, and this fact was described as ``surprising''. The system-environment entanglement in thermal states in the quantum Brownian motion model was further investigated in \cite{Hilt_PRA09}, where it was shown that it disappears above a certain temperature. Qubit-environment entanglement in the case of an environment consisting of non-interacting bosons, coupled linearly to the qubit, was considered in \cite{Pernice_PRA11}, where it was shown that for a qubit initialized in a pure state and an environment in a thermal equilibrium state at finite temperature, decoherence is always accompanied by nonzero QEE - but not necessarily so for
a qubit being initially in a mixed state. 

Clearly, the fact that the issue of correlation between a system's decoherence and the generation of system-environment entanglement is a nontrivial one in a general setting, has been a subject of intense attention.
However, a simple theoretical criterion showing when 
qubit pure dephasing is accompanied by generation of QEE
(for a qubit initialized in a pure state interacting with {\it any} finite-dimensional environment) has been formulated only quite recently \cite{Roszak_PRA15,Roszak_PRA18}: QEE is generated if and only if the evolutions of the environment conditioned on two pointer states of the qubit lead to distinct states of the environment. Obviously such a situation is incompatible with treating the environment as an entity that evolves independently of the qubit. 

In this paper we propose a very simple experimental scheme, which can be used to test the generation of QEE during
the joint evolution of the qubit and its the environment initially in a product state, where the qubit is in a superposition of its pointer states, and the interaction leads to its pure dephasing.
The scheme relies on the fact that only for entangling evolutions the environment behaves in a distinct way depending on which pointer state the qubit is in.
Hence, only if an evolution is entangling can there be a difference in the evolution of qubit coherence when the environment has been allowed to evolve for a finite time in the presence of qubit state $|0\rangle$ or $|1\rangle$ before a qubit superposition state was created. The observation of distinct evolutions of qubit coherence for the two preparation procedures is therefore a QEE witness. Furthermore, the fact that the evolution of the environment does depend on the state of the qubit is incompatible with the assumption that the environment is an entity that evolves independently of the qubit \footnote{Note that this does not necessarily preclude using a classical noise picture for calculation of the dephasing, but one should keep in mind that this is then only a calculation tool.}.
Sensing the capability of the system to generate nonzero QEE during a qubit's dephasing  also proves then that the environmental influence cannot be described as external classical noise.

Critically, unlike the scheme proposed in \cite{Roszak_PRA15}, here
only measurements on the qubit, not on the environment, are required, making the scheme completely straightforward to implement. 
Although it is common knowledge that detection of entanglement between two systems requires, in general, measurements on both of the systems, here we need to measure only one of the systems (the qubit), because the problem is constrained: we are interested in entanglement generated during pure dephasing of the qubit interacting with an environment.
We illustrate the concept with a calculation performed for a nitrogen-vacancy (NV) center in diamond, a spin qubit coupled to a nuclear spin environment that is widely used for noise spectroscopy and nanoscale nuclear magnetic resonance purposes \cite{Degen_RMP17,Wrachtrup_JMR16}.

The paper is organized in the following way. In Sec.~\ref{sec:PD} we discuss 
	the pure dephasing model of decoherence and its applicability.
	In Sec.~\ref{sec:protocol} we describe the protocol for the detection of the system's capacity to generate qubit-environment entanglement which is the central result
	of this paper. The discussion of the significance of this protocol for sensing the non-classical nature of the environmental noise is given in Sec.~\ref{sec:classical}. Then, in Sec.~\ref{sec:NV} we predict the performance of the protocol applied to an the NV center spin qubit interacting with a partially polarized nuclear environment. Sec.~\ref{sec:concl} concludes the paper.

\section{Pure dephasing Hamiltonian} \label{sec:PD}

The system under study is composed of a qubit and an environment of arbitrary size.
The interaction between the two is such that the effect of the environment on
the qubit can only lead to its pure dephasing, so processes which affect the occupations
of the qubit are not allowed. Such a class of Hamiltonians can be simply defined,
since the condition for decoherence to be limited to pure dephasing amounts to
the fact that the free qubit Hamiltonian must commute with the interaction terms.

We choose states $|0\rangle$ and $|1\rangle$ to be qubit pointer states, 
which allows us to write an explicit
general form of the pure-dephasing Hamiltonian,
\begin{equation}
\label{H}
\hat{H}=
\sum_{i=0,1}\varepsilon_{i}|i\rangle\langle i|+\hat{H}_{\mathrm{E}}+ 
\sum_{i=0,1}|i\rangle\langle i|\otimes{\hat{V_i}} \,\, .
\end{equation}
Here the first term describes the free evolution of the qubit
and $\varepsilon_{i}$ 
are the energies of the qubit states, 
the second term describes the environment, 
while the last term describes the qubit-environment (QE) interaction.
The environmental operators $\hat{V_0}$ and $\hat{V_1}$ are arbitrary, the same
as the environment Hamiltonian, $\hat{H}_{\mathrm{E}}$.

QE evolution can be formally solved for Hamiltonians of this class and
the QE evolution operator $\hat{U}(t)\! =\! \exp(-i\hat{H}t)$
can be written as
\begin{equation}
\label{u}
\hat{U}(t) = |0\rangle\langle 0|\otimes\hat{w}_0(t)+ |1\rangle\langle 1|\otimes \hat{w}_1(t),
\end{equation}
where the operators which describe the evolution of the environment
conditional on the state of the qubit are given by
\begin{equation}
\label{wi}
\hat{w}_i(t)= \exp(-i\hat{H}_{i}t),
\end{equation}
with $i=0,1$. The operators $\hat{H}_{i} \! =\! \hat{H}_{\mathrm{E}} + \hat{V}_{i}$
contain the free Hamiltonian of the environment and the appropriate part of the 
interaction.

The above Hamiltonian is not only a paradigmatic model of decoherence (as it describes the simplest setting in which environment causes dephasing of superpositions of pointer states \cite{Zurek_PRD82,Zurek_RMP03,Schlosshauer_book}, but it also describes the dominant decoherence process for most types of currently researched qubits, e.g.~spin qubits in quantum dots \cite{Cywinski_APPA11,Yang_RPP17,Medford_PRL12,Malinowski_PRL17,Malinowski_NN17}, spin qubits based on NV centers \cite{Zhao_PRB12,deLange_Science10,Witzel_PRB12,Yang_RPP17} and electrons bound to donors \cite{Witzel_PRL10}, trapped ions \cite{Biercuk_Nature09,Monz_PRL11}, and exciton-based qubits \cite{Roszak_PLA06,Krzywda_SR16,Salamon_PRA17,Paz_PRA17}. In all these systems the dephasing of a superposition of $\ket{0}$ and $\ket{1}$ state occurs on timescales orders of magnitude shorter than timescale on which energy is exchanged between the qubit and the environment, and consequently populations of these states are modified. 

It is also worth noting that the absence of transverse couplings $\propto \hat{\sigma}_x \hat{V}_x +\hat{\sigma}_y \hat{V}_y$ in the QE Hamiltonian is not necessary for pure dephasing to be the process that limits the coherence of the qubit. It is enough for the energy scale $\Delta E \equiv \epsilon_1 - \epsilon_0$ of the qubit's Hamiltonian to be much larger than the energy scales associated with these transverse terms, i.e.~$\Delta E \! \gg \! [\mathrm{Tr}_{E}( \hat{R}(0) \hat{V}^{2}_{x/y})]^{1/2}$, where $\hat{R}(0)$ is the density matrix of the environment. If the spectral density of the $\hat{V}_{x,y}$ fluctuations of the environment does not overlap very strongly with frequency range around $\Delta E$, the energy exchange with the environment will be weak, and the qubit's quantization axis will be only slightly tilted away from the $z$ direction by $\hat{V}_{x,y}$ terms.  
In this situation, one can use a Schrieffer-Wolff canonical transformation and obtain an effective pure dephasing Hamiltonian containing the terms $\propto \hat{\sigma}_z \hat{V}_{x,y}^2/2\Omega$, for examples see e.g.~\cite{Makhlin_PRL04,Falci_PRL05,Yao_PRB06,Bergli_PRB06,Cywinski_PRB09,Ramon_PRB12,Cywinski_PRA14,Szankowski_QIP15,Malinowski_PRL17}.


\section{Protocol of detection of system's capacity to generate qubit-environment entanglement} \label{sec:protocol}
We begin with the main result of Ref.~\cite{Roszak_PRA15},
which provides a criterion to distinguish between entangling and non-entangling
QE evolutions for pure-dephasing processes. The criterion works only
for product initial states of the qubit and the environment, $\ket{\phi}\bra{\phi}\otimes \hat{R}(0)$.
Additionally
the initial state of the qubit has to be pure $\ket{\phi} \! =\! \alpha|0\rangle +\beta|1\rangle$ (for QEE to be generated in a pure-dephasing process, obviously the qubit has to be in a superposition of its pointer states, hence $\alpha,\beta\neq 0$).
There are no constraints on the initial state of the environment, neither on its size or purity, and it is described by the 
density matrix $\hat{R}(0)$. The criterion states that
QEE is present at time $\tau$ after initialization, if and only if $[\hat{w}_0^{\dagger}(\tau)\hat{w}_1(\tau),\hat{R}(0)]\neq 0$. This
condition can be rewritten \cite{Roszak_PRA18} in the more physically meaningful form
\begin{equation}
\label{warunek}
\hat{w}_0(\tau)\hat{R}(0)\hat{w}^{\dagger}_0(\tau)\neq \hat{w}_1(\tau)\hat{R}(0)\hat{w}_1^{\dagger}(\tau). 
\end{equation}
Here the conditional evolution operators of the environment are given by eq.~(\ref{wi}).
Note that $\hat{w}_i(\tau)\hat{R}(0)\hat{w}_i^{\dagger}(\tau)$, $i=0,1$, is the state  of the environment at time $\tau$ conditional on the qubit
being initialized in state $|i\rangle$. Therefore, if and only if QEE is not generated, the evolution of the environment in the presence of either qubit pointer state will be the same, otherwise, it has to differ.

This condition itself provides a straightforward QEE witness, since any observable on
the environment can be used to test it \cite{Roszak_PRA15}. If the qubit would be prepared
initially in state $|0\rangle$ and the time dependence of an observable
on the environment would be measured,
and then the qubit would be prepared in state $|1\rangle$
(for the same initial state of the environment $\hat{R}(0)$) which would be followed by
measuring the time dependence of the same observable, a discrepancy 
at time $\tau$ between the expectation values of said observable would mean that
if the qubit were initially prepared in any superposition state, it would
be entangled with the environment at time $\tau$. 
Since the result would obviously depend on the choice of observable,
the same expectation value at time $\tau$ is inconclusive
(as conclusive testing would require the full knowledge of the conditional density 
matrices of the environment at time $\tau$).
The problem with such tests of QEE is that it requires measurements to be performed
on the environment which is usually hard to access. 

\begin{figure}[tb]
	\includegraphics[width=1\columnwidth]{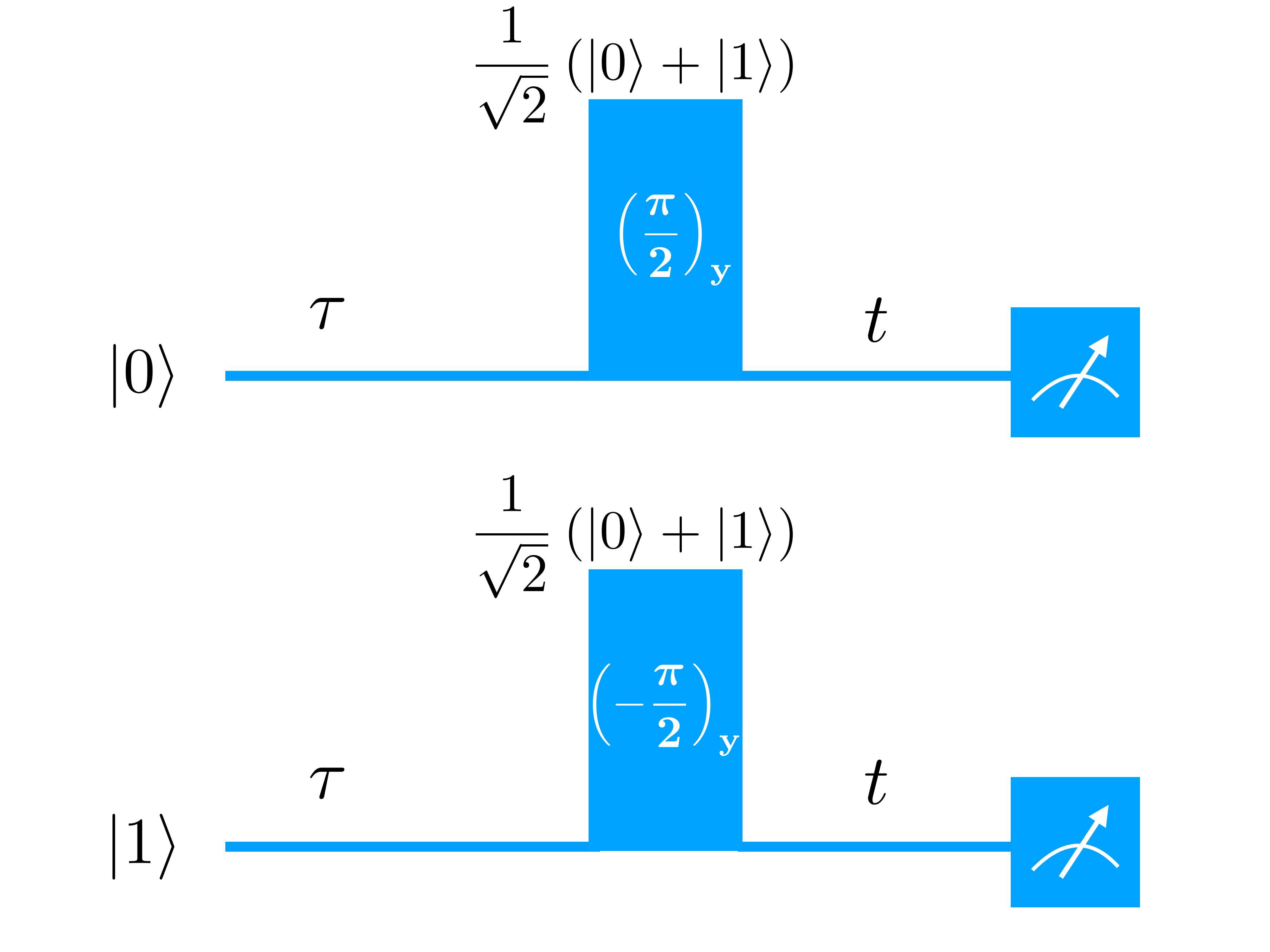}
	\caption{Schematic representation of the protocol for the detection of a system's capacity to generate qubit-environment entanglement. After a preparation time $\tau$ when
		the environment evolves in the presence of state $|0\rangle$ or $|1\rangle$, the
		qubit is (operationally instantaneously) prepared in a superposition state 
		(the same in both cases). Then the evolution of the coherence is measured and results
		in both cases are compared.
	}\label{fig:scheme}
\end{figure}

In the following, we use the fact that QEE generation in the described evolutions
always corresponds to different evolutions of the environment conditional on the
pointer state of the qubit
and propose a scheme for QEE detection, which requires operations and 
measurements on the qubit alone. The protocol is schematically depicted in Fig.~\ref{fig:scheme}.

The idea is to first prepare the qubit in state $|0\rangle$ and let it and the
environment
evolve jointly for time $\tau$. For pure dephasing evolutions this does not change the qubit state, but the environment evolves into $\hat{R}_0(\tau)=\hat{w}_0(\tau)\hat{R}(0)\hat{w}_0^{\dagger}(\tau)$
from its initial state $\hat{R}(0)$. Now, if at time $\tau$ the qubit state is changed to $|\psi\rangle=1/\sqrt{2}\left(|0\rangle+|1\rangle\right)$ by an appropriate unitary operation (the equal superposition state is chosen to maximize the visibility of the effect, but any superposition would work), further evolution will lead to pure dephasing of the qubit and the coherence will evolve according to
 \begin{equation}
 \label{zero}
 \rho_{01}^{(0)}(\tau,t)=\frac{1}{2}\mathrm{Tr}\left(
 \hat{w}_0(t)\hat{w}_0(\tau)\hat{R}(0)
 \hat{w}_0^{\dagger}(\tau)\hat{w}_1^{\dagger}(t)\right),
 \end{equation}
 where $t$ is the time elapsed from time $\tau$. This coherence needs to be measured. Next, if the same procedure is performed with the qubit in state $|1\rangle$ between the initial moment and time $\tau$, the coherence of the superposition qubit state (after time
 $\tau$)
 will evolve according to
 \begin{equation}
 \label{jeden}
 \rho_{01}^{(1)}(\tau,t)=\frac{1}{2}\mathrm{Tr}\left(
 \hat{w}_0(t)\hat{w}_1(\tau)\hat{R}(0)
 \hat{w}_1^{\dagger}(\tau)\hat{w}_1^{\dagger}(t)\right).
 \end{equation}
 Naturally if the separability condition of eq.~(\ref{warunek}) is fulfilled, the evolution given by eq.~(\ref{zero}) would be {\it the same} as the evolution given by eq.~(\ref{jeden}). Otherwise, if at any time $t$, $\rho_{01}^{(0)}(\tau,t)\neq \rho_{01}^{(1)}(\tau,t)$, then there must be QEE at time $\tau$
in a system initially in a product of any qubit superposition state and environmental
state $\hat{R}(0)$. Therefore,
unless the two coherence decay signals, (\ref{zero}) and (\ref{jeden}), are in perfect agreement, pure dephasing of a superposition of qubit states lasting for time $\tau$ must be accompanied by QEE generation.

The scheme outlined above is an entanglement witness, since there
exists one situation when QEE is generated, which it does not detect.
This is the case when $[\hat{w}_0(t_1),\hat{w}_1(t_2)] \! =\! 0$ for all times $t_1$ and $t_2$ (such commutation also implies commutation when one or both operators are Hermitian conjugated), resulting in $\hat{\rho}_{01}^{(0)/(1)}(\tau,t)\! =\! \mathrm{Tr}[\hat{R}(0)\hat{w}_1^\dagger(t)\hat{w}_0(t) ]$. 
This requires $[\hat{H}_0,\hat{H}_1] \! =\! 0$. 
Note that if we do not exactly know the form of $\hat{H}_{E}$ and $\hat{V}_{i}$, we can check if $[\hat{H}_0,\hat{H}_1] \! = \! 0$ by performing a spin-echo experiment, in which a superposition state of the qubit is initialized, it interacts with the environment for time $\tau$, is subjected then to a $\hat{\sigma}_x$ operation, and the coherence  read out  after time $\tau$ elapses again is given by
\begin{equation}
\rho_{01}^{\mathrm{echo}}(\tau,\tau) = \frac{1}{2}\mathrm{Tr}\left(\hat{w}_1(\tau)\hat{w}_0(\tau) \hat{R}(0) 
 \hat{w}_1^{\dagger}(\tau)\hat{w}_0^{\dagger}(\tau) \right) \,\, .
\end{equation}
Perfect recovery of initial coherence for any $\tau$ is thus equivalent to $[\hat{H}_0,\hat{H}_1] \! = \! 0$.

\section{Relation to the classical noise model of the environmental influences} \label{sec:classical}
Let us now connect the above QEE detection scheme with the question of the nature of noise that leads to qubit dephasing. If the dynamics of the environment is completely independent of the presence of the qubit, we can think of it as a source of a {\it field} that evolves in time in some complicated way, essentially stochastic. This field can couple to the two levels of the qubit in a distinct way, so that the Hamiltonian of the qubit exposed to it is
\begin{align}
\hat{H}(t) & = \sum_{i=0,1}\varepsilon_{i}|i\rangle\langle i|+
\sum_{i=0,1}|i\rangle\langle i|\xi_{i}(t)\,\, , \nonumber\\
&  = [\Delta\varepsilon + \Delta \xi(t)]\hat{\sigma}_{z}/2  + [\bar{\varepsilon} + \bar{\xi}(t)]\mathds{1}/2 \,\, ,
\end{align}
where $\xi_{i}$ are stochastic fields coupling to the qubit state $|i\rangle$, $\Delta\varepsilon \!= \! \varepsilon_0 - \varepsilon_1$, $\bar{\varepsilon} \!= \! \varepsilon_0 + \varepsilon_1$, and $\Delta \xi(t)$ and $\bar{\xi}(t)$ are defined in an analogous way. It is now crucial to be aware that the dependence of $\xi_{i}(t)$ on the qubit state $\ket{i}$ {\it does not} mean that the actual dynamics of 
E depends on this state: both $\xi_{i}(t)$ are related to an underlying dynamics of the environment itself, and the dependence on $i$ is due to the fact that the two states might couple to the environmental noise in a distinct way (see below for a simple example).  
The density matrix describing the initialized $|i\rangle$ state of the qubit does not change under the influence of the above Hamiltonian. The evolution for time $\tau$ that precedes the creation of superposition state of the qubit is thus absent, and $\rho^{(0)}_{01}(\tau,t) \! =\! \rho^{(1)}_{01}(\tau,t)$ while being also independent of $\tau$. 
Furthermore, $\bar{\varepsilon}$ and $\bar{\xi}(t)$ drop out from the expression for qubit coherence,
\begin{equation}
\rho_{01}^{(0/1)}(\tau,t) = e^{-i\Delta \varepsilon t} \left\langle e^{-i\int_{0}^{t} \Delta\xi(t')\mathrm{d}t'} \right\rangle \,\, ,
\end{equation}
where $\langle \ldots \rangle$ denotes averaging over realizations of $\Delta \xi(t')$ noise. Therefore, the observation of $\rho_{01}^{(0)}(\tau,t) \! \neq \! \rho_{01}^{(1)}(\tau,t)$ means that the environment {\it cannot} be described as a source of external classical noise acting on the qubit. 

Note that the known result that any pure dephasing evolution of a qubit can be reproduced by replacing the environment by an artificially constructed source of classical noise \cite{Landau_Streater_LAA93,Helm_PRA09,Crow_PRA14}, has no relevance to the above reasoning. We are interested here in making a statement about the  dynamics of the real environment of the given qubit, and the above test allows us to easily notice the situation in which the qubit-environment interaction modifies the dynamics of the environment.

\section{Result for NV center interacting with partially polarized nuclear environment} \label{sec:NV}
We now present an example of a system in which the creation of QEE, and the non-classicality of the environmental noise, can be detected using the scheme described above. We focus on a nitrogen-vacancy (NV) center spin qubit in diamond, which has been a subject of intense research aimed at using it as a nanoscale resolution sensor of magnetic field fluctuations \cite{Staudacher_Science13,DeVience_NN15,Haberle_NN15,Lovchinsky_Science16,Wrachtrup_JMR16,Degen_RMP17}. The low energy degrees of freedom of the NV center constitute an effective electronic spin $S\! = \! 1$, subjected to zero-field splitting $\Delta (S^{z})^2$, with the direction of $z$ axis determined by the geometry of the center. The presence of a finite magnetic field (assumed here to be along the $z$ axis) leads to a splitting of $m_s=\pm 1$ levels, and therefore the energy level spacing is uneven, so that any two-dimensional subspace of the $S\! = \! 1$ manifold can be used as a qubit controlled by microwave electromagnetic fields. We focus on the most widely employed qubit based on $m\! =\! 0$ and $1$ levels. 

The relevant environment of this qubit consists of nuclear spins of either $^{13}C$ spinful isotope naturally present in a diamond lattice, or nuclei of molecules attached to the surface of the diamond crystal \cite{Staudacher_Science13,Lovchinsky_Science16}. Due to a large value of the zero-field splitting ($\Delta \! =\! 2.87$ GHz), and a large ratio of electronic and nuclear gyromagnetic factors, for almost all values of the magnetic field, the energy exchange between the qubit and the environment is very strongly suppressed, and we can safely use the pure dephasing approximation \cite{Zhao_PRB12}. Crucially, the $m\! = \! 0$ state of the qubit is completely decoupled from the nuclear environment (and if the nuclei can be treated as source of classical noise, $\xi_{0}(t)\! =\! 0$ while $\xi_{1}(t)\! \neq \! 0$), so that keeping the qubit in this state between the measurement and re-initialization does not perturb the state of
the environment.
The QE Hamiltonian is thus given by 
\begin{equation}
\hat{H} = (\Delta + \Omega) \ket{1}\bra{1} + \hat{H}_{E} + \ket{1}\bra{1} \otimes\hat{V}_{1} \,\, ,
\end{equation}
where $\Omega \! =\! -\gamma_{e}B_{z}$ with the electron gyromagnetic ratio $\gamma_{e} \! =\!  28.02$ GHz/T, and $\hat{H}_{E} = \sum_{k}\gamma_{n}B_{z}\hat{I}^{z}_{k} + \hat{H}_{\mathrm{nn}}$,  
where $k$ labels the nuclear spins, $\gamma_{n} \! = \! 10.71$ MHz/T for $^{13}$C nuclei, $\hat{I}^{z}_{k}$ is the operator of the $z$ component of the nuclear spin $k$, and $\hat{H}_{\mathrm{nn}}$ contains the internuclear magnetic dipolar interactions. Finally, the interaction term describes the hyperfine NV-nucleus interaction, and
\beq
\hat{V}_{1} = \sum_{k}\sum_{j \in (x,y,z)} \mathbb{A}^{z,j}_k\hat{I}^{j}_{k} \,\, .
\eeq
This coupling contains two parts, the Fermi contact interaction corresponding to non-zero probability of finding an electron bound to an NV center on the location of a given nucleus, and the dipolar coupling. Usually, the former is omitted since the wavefunction of a deep defect is strongly localized and in fact, for an NV center it has non-negligible impact when the distance between the vacancy and the nucleus is not greater than 0.5 nm \cite{Gali2008}.
With both types of interaction, the coupling constants are given by
\begin{equation}
\mathbb{A}^{z,j}_k=\frac{8\pi\gamma_e\gamma_n}{3} \left|\psi_e(\mathbf{r}_k)\right|^2+\frac{\mu_0}{4\pi}\frac{\gamma_e\gamma_n}{r_{k}^3}\left(1-\frac{3(\mathbf{r}_{k}\cdot\hat{\mathbf{j}})(\mathbf{r}_{k}\cdot\hat{\mathbf{z}})}{r_{k}^2}\right) \, ,
\end{equation}
where $\mu_0$ is the magnetic permeability of the vacuum, $\mathbf{r}_{k}$ is a displacement vector between the $k$-th nucleus and the qubit and $\psi_e(\mathbf{r}_k)$ is the wavefunction of an electron from NV center at position of the $k$-th nucleus.

We use $B_z\! =\! 200$ Gauss, which was employed in a few  recent experiments on
 qubit-based characterization of the small groups of nuclei \cite{Staudacher_Science13,DeVience_NN15,Haberle_NN15}. We consider an environment of about 500 spins in a ball of 9 nm radius with the NV at its center.
Using a well-established and systematic procedure of Cluster-Correlation Expansion (CCE) \cite{Yang_CCE_PRB08,Zhao_PRB12,Kwiatkowski_PRB18}, we have checked {that neither increasing the size of the environment, nor including the interactions within the environment}, $\hat{H}_{\mathrm{nn}}$, {gives} any visible contribution to decoherence of a freely evolving qubit. This is because the coherence decays practically completely before {the more remote nuclei can have an appreciable influence on the qubit} and the inter-nuclear correlations created by interactions become significant. We can then focus on single-spin precession as the only source of dynamics within the environment.

We consider a dynamically polarized nuclear environment. The justification is twofold: (1) without dynamic nuclear polarization (DNP), the density operator of the environment at low fields is $\hat{R}(0) \! \propto \! \mathds{1}$, and according to Eq.~(\ref{warunek}) there is no QEE for such initial states; (2) DNP of the environment of an NV center has been recently mastered \cite{London2013,Fischer2013a,Pagliero2018,Wunderlich2017,Alvarez2015,King2015,  Scheuer2017,Poggiali2017,Hovav2018} and its presence is expected to enhance the signal that the qubit experiences. We assume that $\hat{R}(0)$ does not contain any correlations between the nuclei, i.e.~$\hat{R}(0) \! = \! \bigotimes_{k} \hat{\rho}_{k}$, where $\hat{\rho}_{k}$ is the density matrix of $k$-th nucleus, given in the case of spin-$1/2$ nuclei by $\hat{\rho}_k \! =\! \frac{1}{2}(\mathds{1} + p_{k}\hat{I}^{z}_{k})$, where $p_k \! \in \! [-1,1]$ is the polarization of the $k$-th nucleus. Below we show results for the case in which all the
(spinful) nuclei within a radius $r_p$ from the qubit are fully polarized, while the remaining nuclei are in a completely mixed state. This mimics the experimentally relevant situation, in which the DNP is created by appropriate prior manipulations on the qubit, that lead to polarization of nuclei that are most strongly coupled to it.

\begin{figure}[tb]
	\includegraphics[width=0.95\columnwidth]{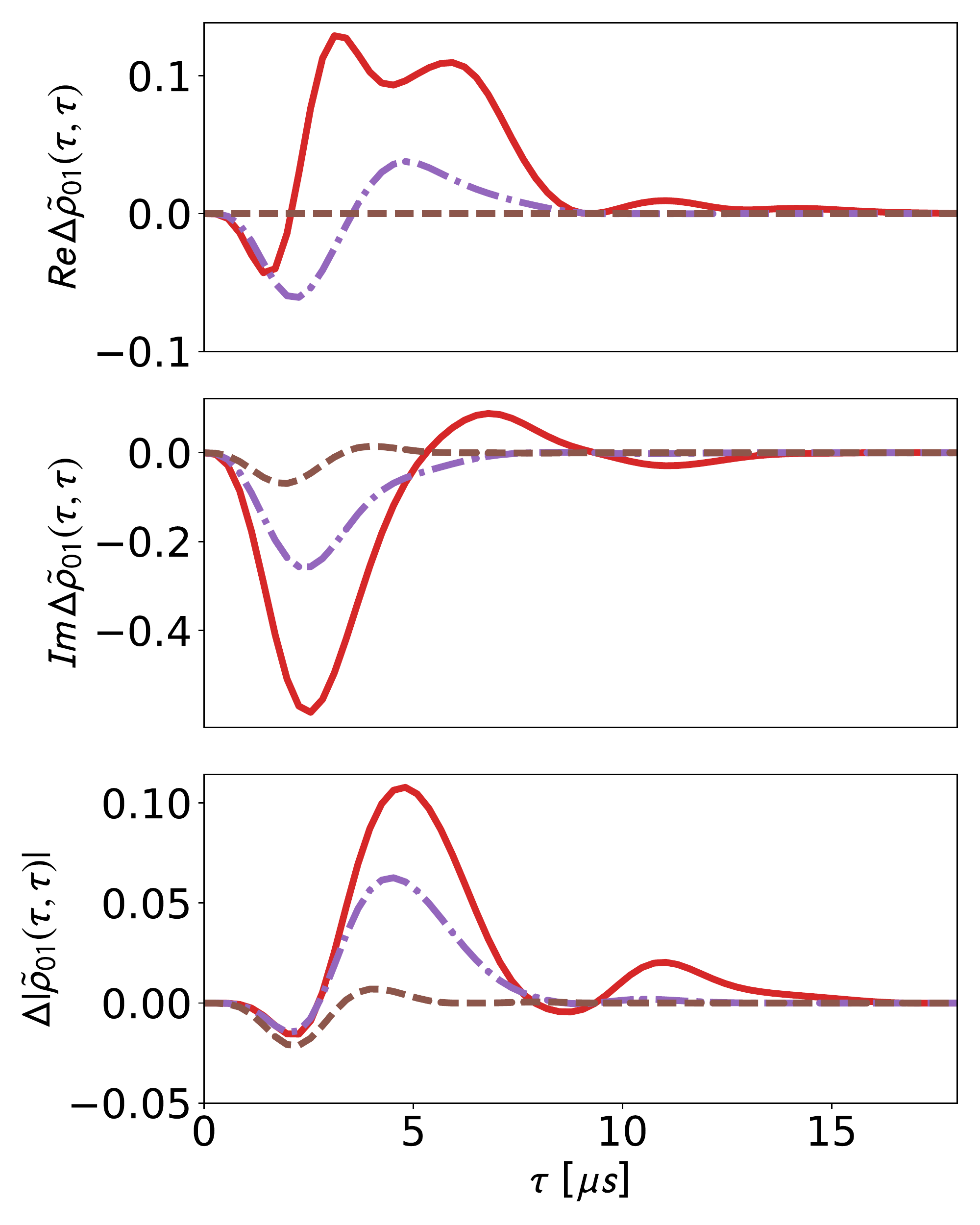}
	\caption{Difference between real (upper panel) and imaginary (middle panel) parts, as well as the absolute value (lower panel) of $\rho^{(0)}_{01}(\tau,t)$ and $\rho^{(1)}_{01}(\tau,t)$ coherence signals (normalized by the maximum qubit coherence) of an NV center qubit interacting with partially polarized nuclear environment
		for a single randomly generated spatial arrangement of
		environmental spins, plotted for $t \! = \!\tau$ at magnetic field $B_z \! =\! 200$ G. 
		Dashed, dot-dashed and solid lines correspond to polarization radius $r_{p} \!= \!$ $0.6$ (one spin polarized), $0.7$ (five spins polarized) and $0.9$ nm
		(nineteen spins polarized), respectively. In the $r_p \! =\! 0.6$ nm case there is only {\it one} polarized spin, hence the lack of evolution in the upper panel, as follows from Eq.~(\ref{eq:DL}).
	}\label{fig:diffRpol}
\end{figure}

\begin{figure}[!h]
	\centering
	\includegraphics[width=1.\columnwidth]{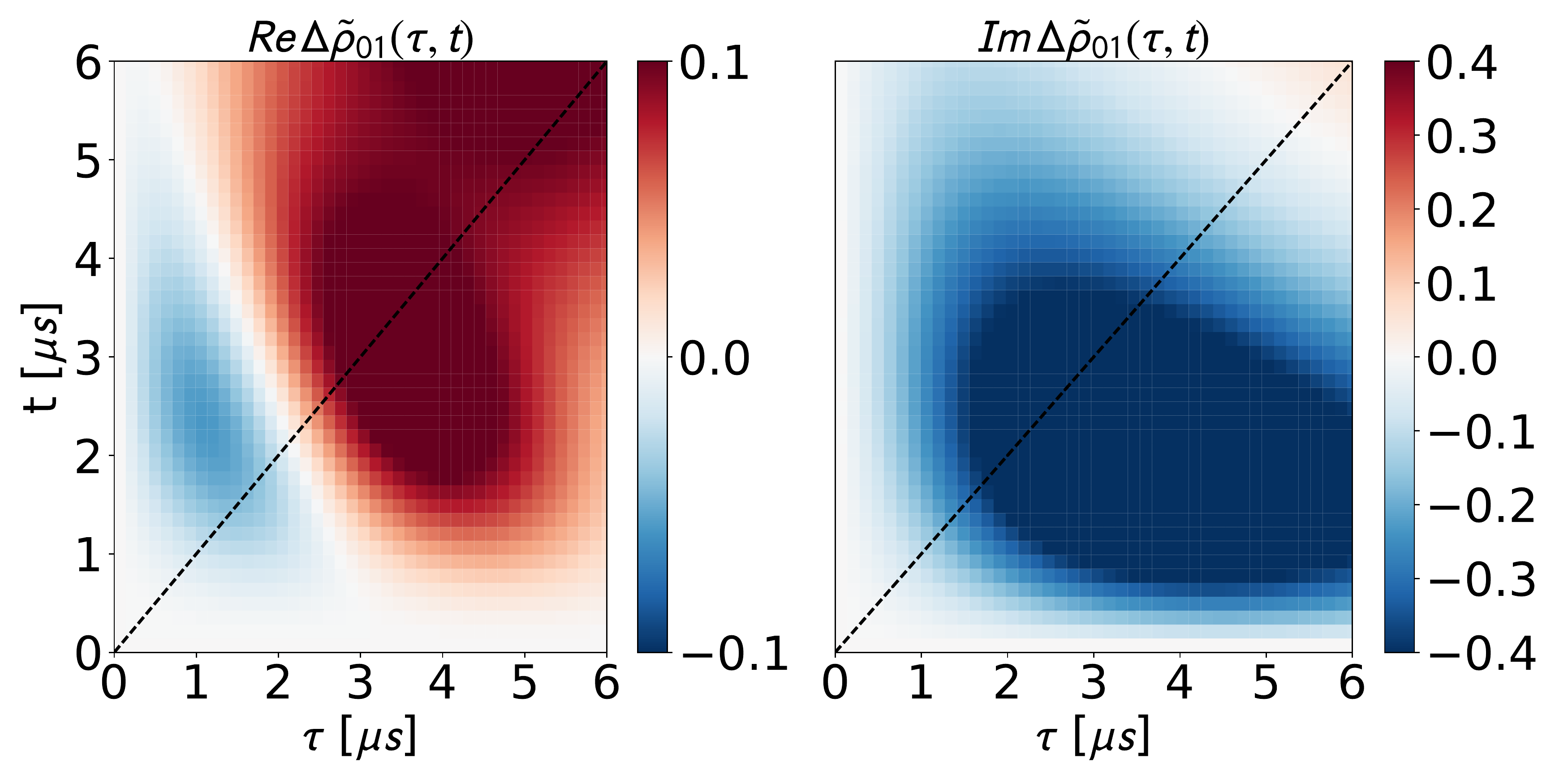}
	\caption{Real (left) and imaginary (right) parts of $\Delta\tilde{\rho}_{01}(\tau,t)$ as functions of $\tau$ and $t$ for all nuclei polarized within radius $r_p \! =\! 0.9$ nm around the qubit at magnetic field $B_z \! =\! 200$ G (spatial arrangement of environmental spins as in Fig.~\ref{fig:diffRpol}). Dashed black line signifies $t=\tau$, which corresponds to the results shown in Fig.~\ref{fig:diffRpol}.
	}\label{fig:2Dmap}
\end{figure}

We work in the rotating frame where the phase accumulated due to the controlled energy splitting of qubit levels is absent. The coherence signal $\rho^{(0/1)}_{01}(\tau,t)$ is then expressed as $\prod_{k} L^{(0/1)}_k(\tau,t)$, where $L^{(0/1)}_k(\tau,t)$ are signals that would be obtained if the environment consisted {\it only} of the $k$-th nuclear spin. While the difference of the two signals, $\Delta \rho_{01} \! \equiv \! \rho^{(0)}_{01} - \rho^{(1)}_{01}$, is not easily expressed through $\Delta L_{k} \! \equiv \! L^{(0)}_{k} - L^{(1)}_{k}$, it is instructive to look at such quantities, which describe the difference of the coherence decay signals for an environment consisting of a single spin:
\beq
\Delta L_{k} = -i\frac{p_k A_x^2\sin\left(\omega_{xz}\frac{t}{2}\right)\sin\left(\omega_{xz}\frac{\tau+t}{2}\right)\sin\left(\omega\frac{\tau}{2}\right)}{\omega_{xz}^2} \,\, , \label{eq:DL}
\eeq
in which for clarity we only kept the $z$ and $x$ couplings to the qubit ($A_{z}\! =\! \mathds{A}^{z,z}_{k}$ and $A_{x}\! =\! \mathds{A}^{z,x}_{k}$), $\omega_{xz} \! \equiv \! \sqrt{A^{2}_{x}+(A_z+\omega)^2}$. The above expression vanishes when either $p_k\! =\! 0$, $A_{x}\!= \! 0$, or  $\tau \! =\! 0$. Since $\Delta L_k$ is purely imaginary, we should carefully inspect both real and imaginary parts of $\Delta \rho_{01}$, not just its magnitude. 
\begin{figure*}[ht]
\subfigure{\includegraphics[trim={0.cm 0.5cm 0cm 0.0cm},clip,width=0.8\textwidth]{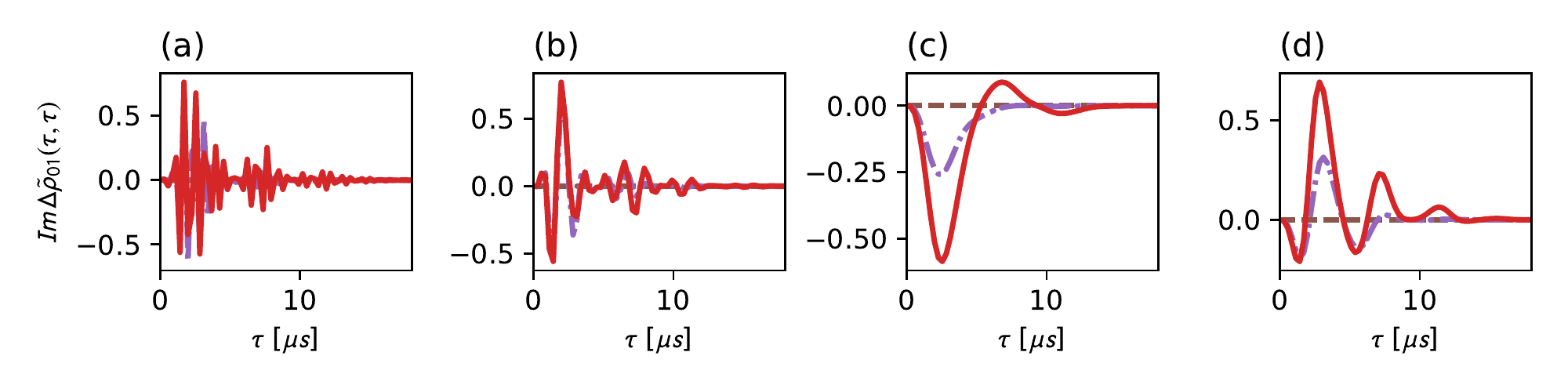}}
\subfigure{\includegraphics[trim={0.cm 0cm 0cm 0.52cm},clip,width=0.8\textwidth]{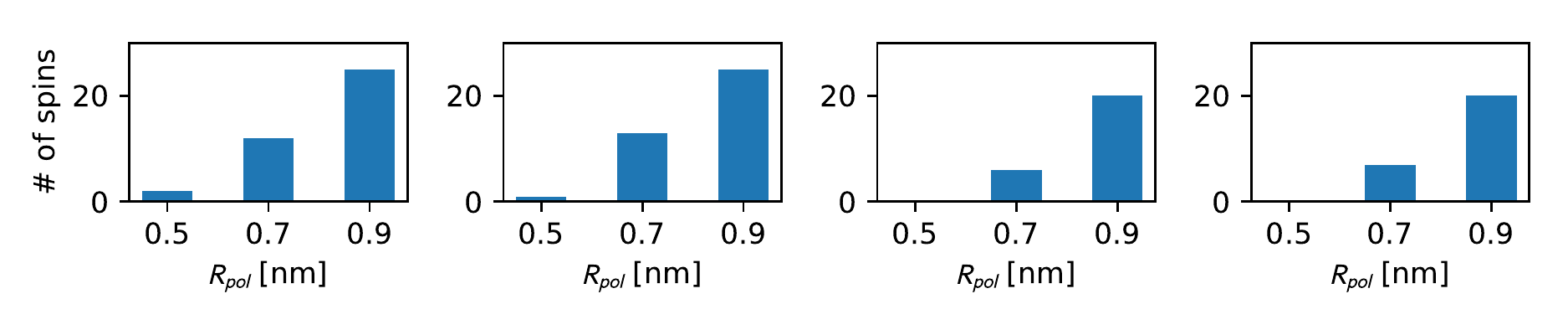}}
\subfigure{\includegraphics[trim={0.cm 0.5cm 0cm 0.4cm},clip,width=0.8\textwidth]{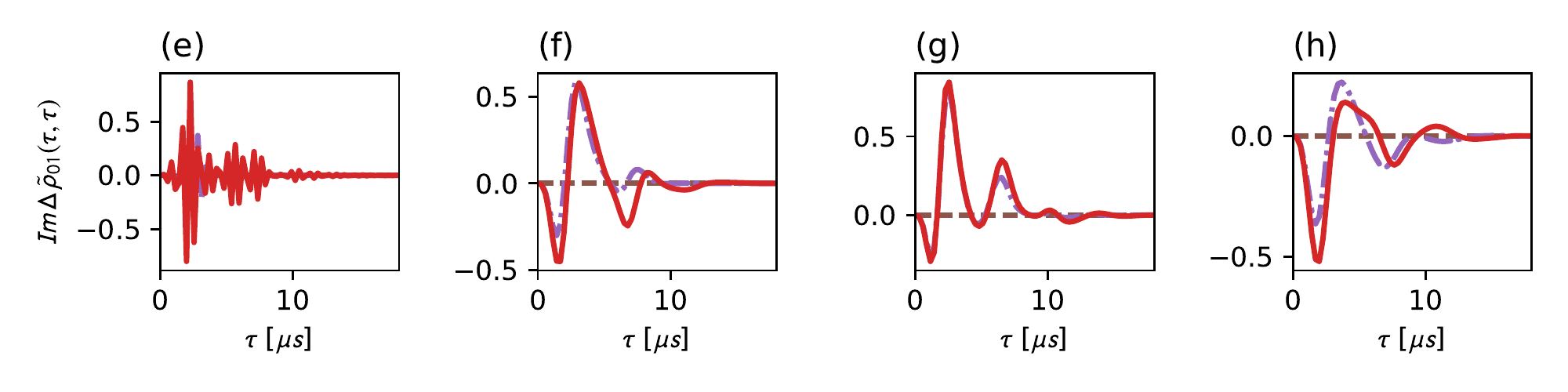}}
\subfigure{\includegraphics[trim={0.cm 0cm 0cm 0.52cm},clip,width=0.8\textwidth]{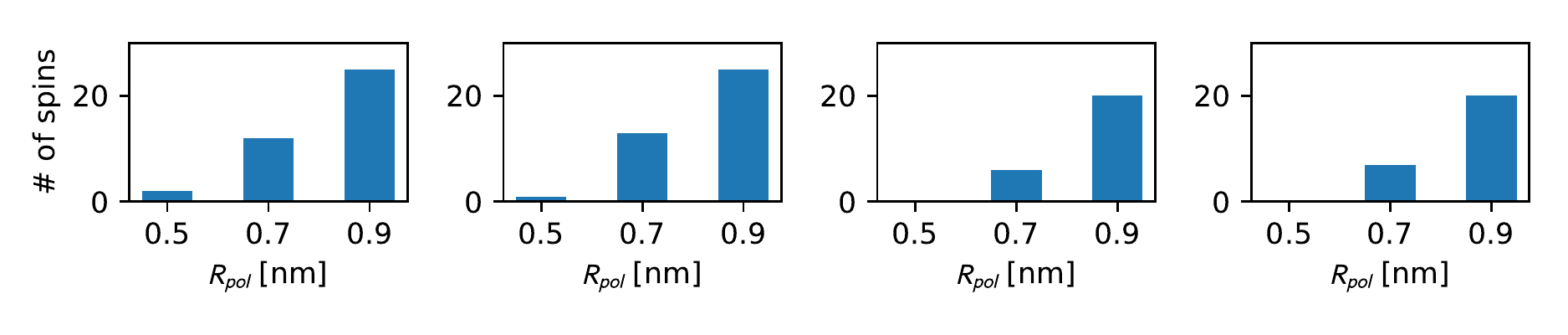}}
	\caption{Imaginary part of $\Delta \tilde{\rho}_{01}(\tau,\tau)$
		of an NV center qubit for 8 different random realizations of the nuclear environment
		at magnetic field $B_z = 200$ G. Dashed, dot-dashed, and solid lines correspond to polarization radius $r_{pol}$ = 0.5, 0.7 and 0.9 nm, respectively.
		The bar graphs show the number of $^{13}$C nuclei corresponding to a given
		$r_{pol}$ in the figure directly above.}
	\label{fig:morereals}
\end{figure*}
The results for an NV center interacting with natural concentration environment of $^{13}C$ spins in diamond, obtained for a single randomly generated spatial arrangement of these spins, are shown in Figs \ref{fig:diffRpol} and \ref{fig:2Dmap}. Both figures show the real and imaginary parts of $\Delta \tilde{\rho}_{01}(\tau,t)=\Delta \rho_{01}(\tau,t)/\rho_{01}(\tau,0)$ (the difference is normalized
by the initial qubit coherence), while Fig.~\ref{fig:diffRpol} additionally contains plots of the difference
between the absolute values of $\rho^{(0)}_{01}(\tau,t)$ and $\rho^{(1)}_{01}(\tau,t)$ (identically normalized).  When for a given delay
time $\tau$ any of these values is
non-zero, this signifies that QEE would be present at time $\tau$
during the joint evolution of an initial product state of any superposition
of the qubit and state $\hat{R}(0)$ of the environment. 
In Fig.~\ref{fig:diffRpol} the results shown
are for equal evolution and delay times, $t\! =\! \tau$, showing that QEE is
present for an initial superposition qubit state throughout the evolution.

Fig.~\ref{fig:morereals} contains the plots of the imaginary part of $\Delta \tilde{\rho}_{01}(\tau,\tau)$ for eight different random realizations of the nuclear
environment supplemented by bar graphs illustrating the number of spinful nuclei
for a given polarization radius. For half of the presented realizations of the environment, there are no $^{13}$C nuclei in the region (a ball of radius 0.5 nm around the vacancy), where the Fermi contact part has to be taken into account.
In fact, there is roughly a 45\% probability to find a realization containing a $^{13} C$  nucleus whose Fermi contact coupling should be included. For such realizations,
the Fermi contact coupling affects the coupling parallel to the quantization axes, which, according to Eq. \eqref{eq:DL}, modifies denominator of the expression, but also produces much faster oscillations in difference of real as well as imaginary parts of coherence.

Results shown in Fig. \ref{fig:morereals} provide additional evidence for the high magnitude of the QEE signal.
The difference in both real and imaginary  parts (corresponding to measurement of $\hat{\sigma}_{x}$ and $\hat{\sigma}_{y}$ of the qubit initialized in $(\ket{0}+\ket{1})/\sqrt{2}$ state) can reach $40$\%
with respect to the initial coherence. Hence, the detection of QEE with the current level of control and readout quality in NV center qubits should be possible.

\section{Discussion and conclusion \label{sec:concl}}
In conclusion, we have described a simple experimental protocol that allows to check, if QEE generation accompanies pure dephasing of a qubit. Importantly, this protocol requires operations to be performed only on the qubit.
Although we focus on the fact that the proposed method allows for straightforward experimental verification, it is relevant to note that it is also a good theoretical tool. The advantage over the method of Ref.~\cite{Roszak_PRA15} stems from the fact that only the evolution of qubit coherence (for two different initial states of the environment) needs to be calculated and neither the whole QE state (as in general methods) nor operators acting on the environment have to be found. 
A positive result of such a test not only certifies that QEE is created, but also that the influence of the environment cannot be described as classical (i.e.~independent of the existence of the qubit) noise of either Gaussian or non-Gaussian statistics (note that some tests \cite{Zhao_PRL11} aimed at detecting the non-classical nature of environmental noise were in fact detecting the non-Gaussian statistics of it).
  
We have presented theoretical results of the working of this protocol for an NV center coupled to a partially polarized environment consisting of nuclear spins. {We have predicted a signal of magnitude comparable to the observed coherence, clearly showing that the protocol is robust to single-qubit control errors that in principle could depend on the state of the environment.}
While quantifying the relation between the degree in which the zero-entanglement condition is broken and the magnitude of the signal observed in our protocol is beyond the scope of this paper, the fact that the signal is clearly visible means that the classical picture of environmental noise, while being widely adopted for analysis of data obtained with NV centers coupled to nanoscale nuclear environments, is definitely not exact in the case of the NV center interacting with polarized nuclei. 

\section*{Acknowledgements}
This work is supported by funds of Polish National Science Center (NCN), grant no.~DEC-2015/19/B/ST3/03152.


\begin{thebibliography}{70}%
	\makeatletter
	\providecommand \@ifxundefined [1]{%
		\@ifx{#1\undefined}
	}%
	\providecommand \@ifnum [1]{%
		\ifnum #1\expandafter \@firstoftwo
		\else \expandafter \@secondoftwo
		\fi
	}%
	\providecommand \@ifx [1]{%
		\ifx #1\expandafter \@firstoftwo
		\else \expandafter \@secondoftwo
		\fi
	}%
	\providecommand \natexlab [1]{#1}%
	\providecommand \enquote  [1]{``#1''}%
	\providecommand \bibnamefont  [1]{#1}%
	\providecommand \bibfnamefont [1]{#1}%
	\providecommand \citenamefont [1]{#1}%
	\providecommand \href@noop [0]{\@secondoftwo}%
	\providecommand \href [0]{\begingroup \@sanitize@url \@href}%
	\providecommand \@href[1]{\@@startlink{#1}\@@href}%
	\providecommand \@@href[1]{\endgroup#1\@@endlink}%
	\providecommand \@sanitize@url [0]{\catcode `\\12\catcode `\$12\catcode
		`\&12\catcode `\#12\catcode `\^12\catcode `\_12\catcode `\%12\relax}%
	\providecommand \@@startlink[1]{}%
	\providecommand \@@endlink[0]{}%
	\providecommand \url  [0]{\begingroup\@sanitize@url \@url }%
	\providecommand \@url [1]{\endgroup\@href {#1}{\urlprefix }}%
	\providecommand \urlprefix  [0]{URL }%
	\providecommand \Eprint [0]{\href }%
	\providecommand \doibase [0]{http://dx.doi.org/}%
	\providecommand \selectlanguage [0]{\@gobble}%
	\providecommand \bibinfo  [0]{\@secondoftwo}%
	\providecommand \bibfield  [0]{\@secondoftwo}%
	\providecommand \translation [1]{[#1]}%
	\providecommand \BibitemOpen [0]{}%
	\providecommand \bibitemStop [0]{}%
	\providecommand \bibitemNoStop [0]{.\EOS\space}%
	\providecommand \EOS [0]{\spacefactor3000\relax}%
	\providecommand \BibitemShut  [1]{\csname bibitem#1\endcsname}%
	\let\auto@bib@innerbib\@empty
	\bibitem [{\citenamefont {{\.Z}urek}(2003)}]{Zurek_RMP03}%
	\BibitemOpen
	\bibfield  {author} {\bibinfo {author} {\bibfnamefont {W.~H.}\ \bibnamefont
			{{\.Z}urek}},\ }\href {\doibase 10.1103/RevModPhys.75.715} {\bibfield
		{journal} {\bibinfo  {journal} {Rev.\ Mod.\ Phys.}\ }\textbf {\bibinfo
			{volume} {75}},\ \bibinfo {pages} {715} (\bibinfo {year} {2003})}\BibitemShut
	{NoStop}%
	\bibitem [{\citenamefont {Hornberger}(2009)}]{Hornberger_LNP09}%
	\BibitemOpen
	\bibfield  {author} {\bibinfo {author} {\bibfnamefont {K.}~\bibnamefont
			{Hornberger}},\ }\href {\doibase 10.1007/978-3-540-88169-8_5} {\bibfield
		{journal} {\bibinfo  {journal} {Lect.~Notes Phys.}\ }\textbf {\bibinfo
			{volume} {768}},\ \bibinfo {pages} {221} (\bibinfo {year}
		{2009})}\BibitemShut {NoStop}%
	\bibitem [{\citenamefont {{\.Z}urek}(1982)}]{Zurek_PRD82}%
	\BibitemOpen
	\bibfield  {author} {\bibinfo {author} {\bibfnamefont {W.~H.}\ \bibnamefont
			{{\.Z}urek}},\ }\href {\doibase 10.1103/PhysRevD.26.1862} {\bibfield
		{journal} {\bibinfo  {journal} {Phys. Rev. D}\ }\textbf {\bibinfo {volume}
			{26}},\ \bibinfo {pages} {1862} (\bibinfo {year} {1982})}\BibitemShut
	{NoStop}%
	\bibitem [{\citenamefont {Schlosshauer}(2007)}]{Schlosshauer_book}%
	\BibitemOpen
	\bibfield  {author} {\bibinfo {author} {\bibfnamefont {M.}~\bibnamefont
			{Schlosshauer}},\ }\href@noop {} {\emph {\bibinfo {title} {Decoherence and
				the Quantum-to-Classical Transition}}}\ (\bibinfo  {publisher} {Springer},\
	\bibinfo {address} {Berlin/Heidelberg},\ \bibinfo {year} {2007})\BibitemShut
	{NoStop}%
	\bibitem [{\citenamefont {Zurek}(2003)}]{Zurek_revisited}%
	\BibitemOpen
	\bibfield  {author} {\bibinfo {author} {\bibfnamefont {W.~H.}\ \bibnamefont
			{Zurek}},\ }\href {http://arxiv.org/abs/quant-ph/0306072} {\bibfield
		{journal} {\bibinfo  {journal} {arXiv:quant-ph/0306072}\ } (\bibinfo {year}
		{2003})}\BibitemShut {NoStop}%
	\bibitem [{\citenamefont {Degen}\ \emph {et~al.}(2017)\citenamefont {Degen},
		\citenamefont {Reinhard},\ and\ \citenamefont {Cappellaro}}]{Degen_RMP17}%
	\BibitemOpen
	\bibfield  {author} {\bibinfo {author} {\bibfnamefont {C.~L.}\ \bibnamefont
			{Degen}}, \bibinfo {author} {\bibfnamefont {F.}~\bibnamefont {Reinhard}}, \
		and\ \bibinfo {author} {\bibfnamefont {P.}~\bibnamefont {Cappellaro}},\
	}\href {\doibase 10.1103/RevModPhys.89.035002} {\bibfield  {journal}
		{\bibinfo  {journal} {Rev. Mod. Phys.}\ }\textbf {\bibinfo {volume} {89}},\
		\bibinfo {pages} {035002} (\bibinfo {year} {2017})}\BibitemShut {NoStop}%
	\bibitem [{\citenamefont {Sza\'nkowski}\ \emph {et~al.}(2017)\citenamefont
		{Sza\'nkowski}, \citenamefont {Ramon}, \citenamefont {Krzywda}, \citenamefont
		{Kwiatkowski},\ and\ \citenamefont {Cywi\'nski}}]{Szankowski_JPCM17}%
	\BibitemOpen
	\bibfield  {author} {\bibinfo {author} {\bibfnamefont {P.}~\bibnamefont
			{Sza\'nkowski}}, \bibinfo {author} {\bibfnamefont {G.}~\bibnamefont {Ramon}},
		\bibinfo {author} {\bibfnamefont {J.}~\bibnamefont {Krzywda}}, \bibinfo
		{author} {\bibfnamefont {D.}~\bibnamefont {Kwiatkowski}}, \ and\ \bibinfo
		{author} {\bibfnamefont {{\L}.}~\bibnamefont {Cywi\'nski}},\ }\href {\doibase
		10.1088/1361-648X/aa7648} {\bibfield  {journal} {\bibinfo  {journal} {J.
				Phys.:Condens. Matter}\ }\textbf {\bibinfo {volume} {29}},\ \bibinfo {pages}
		{333001} (\bibinfo {year} {2017})}\BibitemShut {NoStop}%
	\bibitem [{\citenamefont {Nakamura}\ \emph {et~al.}(2002)\citenamefont
		{Nakamura}, \citenamefont {Pashkin}, \citenamefont {Yamamoto},\ and\
		\citenamefont {Tsai}}]{Nakamura_PRL02}%
	\BibitemOpen
	\bibfield  {author} {\bibinfo {author} {\bibfnamefont {Y.}~\bibnamefont
			{Nakamura}}, \bibinfo {author} {\bibfnamefont {Y.~A.}\ \bibnamefont
			{Pashkin}}, \bibinfo {author} {\bibfnamefont {T.}~\bibnamefont {Yamamoto}}, \
		and\ \bibinfo {author} {\bibfnamefont {J.~S.}\ \bibnamefont {Tsai}},\
	}\href@noop {} {\bibfield  {journal} {\bibinfo  {journal} {Phys.\ Rev.\
				Lett.}\ }\textbf {\bibinfo {volume} {88}},\ \bibinfo {pages} {047901}
		(\bibinfo {year} {2002})}\BibitemShut {NoStop}%
	\bibitem [{\citenamefont {Roszak}\ and\ \citenamefont
		{Machnikowski}(2006{\natexlab{a}})}]{Roszak_PRA06}%
	\BibitemOpen
	\bibfield  {author} {\bibinfo {author} {\bibfnamefont {K.}~\bibnamefont
			{Roszak}}\ and\ \bibinfo {author} {\bibfnamefont {P.}~\bibnamefont
			{Machnikowski}},\ }\href {\doibase 10.1103/PhysRevA.73.022313} {\bibfield
		{journal} {\bibinfo  {journal} {Phys. Rev. A}\ }\textbf {\bibinfo {volume}
			{73}},\ \bibinfo {pages} {022313} (\bibinfo {year}
		{2006}{\natexlab{a}})}\BibitemShut {NoStop}%
	\bibitem [{\citenamefont {Biercuk}\ \emph {et~al.}(2009)\citenamefont
		{Biercuk}, \citenamefont {Uys}, \citenamefont {VanDevender}, \citenamefont
		{Shiga}, \citenamefont {Itano},\ and\ \citenamefont
		{Bollinger}}]{Biercuk_Nature09}%
	\BibitemOpen
	\bibfield  {author} {\bibinfo {author} {\bibfnamefont {M.~J.}\ \bibnamefont
			{Biercuk}}, \bibinfo {author} {\bibfnamefont {H.}~\bibnamefont {Uys}},
		\bibinfo {author} {\bibfnamefont {A.~P.}\ \bibnamefont {VanDevender}},
		\bibinfo {author} {\bibfnamefont {N.}~\bibnamefont {Shiga}}, \bibinfo
		{author} {\bibfnamefont {W.~M.}\ \bibnamefont {Itano}}, \ and\ \bibinfo
		{author} {\bibfnamefont {J.~J.}\ \bibnamefont {Bollinger}},\ }\href {\doibase
		10.1038/nature07951} {\bibfield  {journal} {\bibinfo  {journal} {Nature}\
		}\textbf {\bibinfo {volume} {458}},\ \bibinfo {pages} {996} (\bibinfo {year}
		{2009})}\BibitemShut {NoStop}%
	\bibitem [{\citenamefont {Bylander}\ \emph {et~al.}(2011)\citenamefont
		{Bylander}, \citenamefont {Gustavsson}, \citenamefont {Yan}, \citenamefont
		{Yoshihara}, \citenamefont {Harrabi}, \citenamefont {Fitch}, \citenamefont
		{Cory}, \citenamefont {Nakamura}, \citenamefont {Tsai},\ and\ \citenamefont
		{Oliver}}]{Bylander_NP11}%
	\BibitemOpen
	\bibfield  {author} {\bibinfo {author} {\bibfnamefont {J.}~\bibnamefont
			{Bylander}}, \bibinfo {author} {\bibfnamefont {S.}~\bibnamefont
			{Gustavsson}}, \bibinfo {author} {\bibfnamefont {F.}~\bibnamefont {Yan}},
		\bibinfo {author} {\bibfnamefont {F.}~\bibnamefont {Yoshihara}}, \bibinfo
		{author} {\bibfnamefont {K.}~\bibnamefont {Harrabi}}, \bibinfo {author}
		{\bibfnamefont {G.}~\bibnamefont {Fitch}}, \bibinfo {author} {\bibfnamefont
			{D.~G.}\ \bibnamefont {Cory}}, \bibinfo {author} {\bibfnamefont
			{Y.}~\bibnamefont {Nakamura}}, \bibinfo {author} {\bibfnamefont {J.-S.}\
			\bibnamefont {Tsai}}, \ and\ \bibinfo {author} {\bibfnamefont {W.~D.}\
			\bibnamefont {Oliver}},\ }\href {\doibase 10.1038/nphys1994} {\bibfield
		{journal} {\bibinfo  {journal} {Nat.~Phys.}\ }\textbf {\bibinfo {volume}
			{7}},\ \bibinfo {pages} {565} (\bibinfo {year} {2011})}\BibitemShut {NoStop}%
	\bibitem [{\citenamefont {Medford}\ \emph {et~al.}(2012)\citenamefont
		{Medford}, \citenamefont {Cywi\'{n}ski}, \citenamefont {Barthel},
		\citenamefont {Marcus}, \citenamefont {Hanson},\ and\ \citenamefont
		{Gossard}}]{Medford_PRL12}%
	\BibitemOpen
	\bibfield  {author} {\bibinfo {author} {\bibfnamefont {J.}~\bibnamefont
			{Medford}}, \bibinfo {author} {\bibfnamefont {{\L}.}~\bibnamefont
			{Cywi\'{n}ski}}, \bibinfo {author} {\bibfnamefont {C.}~\bibnamefont
			{Barthel}}, \bibinfo {author} {\bibfnamefont {C.~M.}\ \bibnamefont {Marcus}},
		\bibinfo {author} {\bibfnamefont {M.~P.}\ \bibnamefont {Hanson}}, \ and\
		\bibinfo {author} {\bibfnamefont {A.~C.}\ \bibnamefont {Gossard}},\ }\href
	{\doibase 10.1103/PhysRevLett.108.086802} {\bibfield  {journal} {\bibinfo
			{journal} {Phys.\ Rev.\ Lett.}\ }\textbf {\bibinfo {volume} {108}},\ \bibinfo
		{pages} {086802} (\bibinfo {year} {2012})}\BibitemShut {NoStop}%
	\bibitem [{\citenamefont {Staudacher}\ \emph {et~al.}(2013)\citenamefont
		{Staudacher}, \citenamefont {Shi}, \citenamefont {Pezzagna}, \citenamefont
		{Meijer}, \citenamefont {Du}, \citenamefont {Meriles}, \citenamefont
		{Reinhard},\ and\ \citenamefont {Wrachtrup}}]{Staudacher_Science13}%
	\BibitemOpen
	\bibfield  {author} {\bibinfo {author} {\bibfnamefont {T.}~\bibnamefont
			{Staudacher}}, \bibinfo {author} {\bibfnamefont {F.}~\bibnamefont {Shi}},
		\bibinfo {author} {\bibfnamefont {S.}~\bibnamefont {Pezzagna}}, \bibinfo
		{author} {\bibfnamefont {J.}~\bibnamefont {Meijer}}, \bibinfo {author}
		{\bibfnamefont {J.}~\bibnamefont {Du}}, \bibinfo {author} {\bibfnamefont
			{C.~A.}\ \bibnamefont {Meriles}}, \bibinfo {author} {\bibfnamefont
			{F.}~\bibnamefont {Reinhard}}, \ and\ \bibinfo {author} {\bibfnamefont
			{J.}~\bibnamefont {Wrachtrup}},\ }\href {\doibase 10.1126/science.1231675}
	{\bibfield  {journal} {\bibinfo  {journal} {Science}\ }\textbf {\bibinfo
			{volume} {339}},\ \bibinfo {pages} {561} (\bibinfo {year}
		{2013})}\BibitemShut {NoStop}%
	\bibitem [{\citenamefont {Muhonen}\ \emph {et~al.}(2014)\citenamefont
		{Muhonen}, \citenamefont {Dehollain}, \citenamefont {Laucht}, \citenamefont
		{Hudson}, \citenamefont {Kalra}, \citenamefont {Sekiguchi}, \citenamefont
		{Itoh}, \citenamefont {Jamieson}, \citenamefont {McCallum}, \citenamefont
		{Dzurak},\ and\ \citenamefont {Morello}}]{Muhonen_NN14}%
	\BibitemOpen
	\bibfield  {author} {\bibinfo {author} {\bibfnamefont {J.~T.}\ \bibnamefont
			{Muhonen}}, \bibinfo {author} {\bibfnamefont {J.~P.}\ \bibnamefont
			{Dehollain}}, \bibinfo {author} {\bibfnamefont {A.}~\bibnamefont {Laucht}},
		\bibinfo {author} {\bibfnamefont {F.~E.}\ \bibnamefont {Hudson}}, \bibinfo
		{author} {\bibfnamefont {R.}~\bibnamefont {Kalra}}, \bibinfo {author}
		{\bibfnamefont {T.}~\bibnamefont {Sekiguchi}}, \bibinfo {author}
		{\bibfnamefont {K.~M.}\ \bibnamefont {Itoh}}, \bibinfo {author}
		{\bibfnamefont {D.~N.}\ \bibnamefont {Jamieson}}, \bibinfo {author}
		{\bibfnamefont {J.~C.}\ \bibnamefont {McCallum}}, \bibinfo {author}
		{\bibfnamefont {A.~S.}\ \bibnamefont {Dzurak}}, \ and\ \bibinfo {author}
		{\bibfnamefont {A.}~\bibnamefont {Morello}},\ }\href {\doibase
		10.1038/nnano.2014.211} {\bibfield  {journal} {\bibinfo  {journal} {Nature
				Nanotechnology}\ }\textbf {\bibinfo {volume} {9}},\ \bibinfo {pages} {986}
		(\bibinfo {year} {2014})}\BibitemShut {NoStop}%
	\bibitem [{\citenamefont {Malinowski}\ \emph
		{et~al.}(2017{\natexlab{a}})\citenamefont {Malinowski}, \citenamefont
		{Martins}, \citenamefont {Cywi{\'n}ski}, \citenamefont {Rudner},
		\citenamefont {Nissen}, \citenamefont {Fallahi}, \citenamefont {Gardner},
		\citenamefont {Manfra}, \citenamefont {Marcus},\ and\ \citenamefont
		{Kuemmeth}}]{Malinowski_PRL17}%
	\BibitemOpen
	\bibfield  {author} {\bibinfo {author} {\bibfnamefont {F.~K.}\ \bibnamefont
			{Malinowski}}, \bibinfo {author} {\bibfnamefont {F.}~\bibnamefont {Martins}},
		\bibinfo {author} {\bibfnamefont {{\L}.}~\bibnamefont {Cywi{\'n}ski}},
		\bibinfo {author} {\bibfnamefont {M.~S.}\ \bibnamefont {Rudner}}, \bibinfo
		{author} {\bibfnamefont {P.~D.}\ \bibnamefont {Nissen}}, \bibinfo {author}
		{\bibfnamefont {S.}~\bibnamefont {Fallahi}}, \bibinfo {author} {\bibfnamefont
			{G.~C.}\ \bibnamefont {Gardner}}, \bibinfo {author} {\bibfnamefont {M.~J.}\
			\bibnamefont {Manfra}}, \bibinfo {author} {\bibfnamefont {C.~M.}\
			\bibnamefont {Marcus}}, \ and\ \bibinfo {author} {\bibfnamefont
			{F.}~\bibnamefont {Kuemmeth}},\ }\href {\doibase
		10.1103/PhysRevLett.118.177702} {\bibfield  {journal} {\bibinfo  {journal}
			{Phys.\ Rev.\ Lett.}\ }\textbf {\bibinfo {volume} {118}},\ \bibinfo {pages}
		{177702} (\bibinfo {year} {2017}{\natexlab{a}})}\BibitemShut {NoStop}%
	\bibitem [{\citenamefont {Norris}\ \emph {et~al.}(2016)\citenamefont {Norris},
		\citenamefont {Paz-Silva},\ and\ \citenamefont {Viola}}]{Norris_PRL16}%
	\BibitemOpen
	\bibfield  {author} {\bibinfo {author} {\bibfnamefont {L.~M.}\ \bibnamefont
			{Norris}}, \bibinfo {author} {\bibfnamefont {G.~A.}\ \bibnamefont
			{Paz-Silva}}, \ and\ \bibinfo {author} {\bibfnamefont {L.}~\bibnamefont
			{Viola}},\ }\href {\doibase 10.1103/PhysRevLett.116.150503} {\bibfield
		{journal} {\bibinfo  {journal} {Phys. Rev. Lett.}\ }\textbf {\bibinfo
			{volume} {116}},\ \bibinfo {pages} {150503} (\bibinfo {year}
		{2016})}\BibitemShut {NoStop}%
	\bibitem [{\citenamefont {Helm}\ and\ \citenamefont
		{Strunz}(2009)}]{Helm_PRA09}%
	\BibitemOpen
	\bibfield  {author} {\bibinfo {author} {\bibfnamefont {J.}~\bibnamefont
			{Helm}}\ and\ \bibinfo {author} {\bibfnamefont {W.~T.}\ \bibnamefont
			{Strunz}},\ }\href {\doibase 10.1103/PhysRevA.80.042108} {\bibfield
		{journal} {\bibinfo  {journal} {Phys. Rev. A}\ }\textbf {\bibinfo {volume}
			{80}},\ \bibinfo {pages} {042108} (\bibinfo {year} {2009})}\BibitemShut
	{NoStop}%
	\bibitem [{\citenamefont {Landau}\ and\ \citenamefont
		{Streater}(1993)}]{Landau_Streater_LAA93}%
	\BibitemOpen
	\bibfield  {author} {\bibinfo {author} {\bibfnamefont {L.~J.}\ \bibnamefont
			{Landau}}\ and\ \bibinfo {author} {\bibfnamefont {R.~F.}\ \bibnamefont
			{Streater}},\ }\href {\doibase 10.1016/0024-3795(93)90274-R} {\bibfield
		{journal} {\bibinfo  {journal} {Linear Algebr. Appl.}\ }\textbf {\bibinfo
			{volume} {193}},\ \bibinfo {pages} {107} (\bibinfo {year}
		{1993})}\BibitemShut {NoStop}%
	\bibitem [{\citenamefont {Crow}\ and\ \citenamefont
		{Joynt}(2014)}]{Crow_PRA14}%
	\BibitemOpen
	\bibfield  {author} {\bibinfo {author} {\bibfnamefont {D.}~\bibnamefont
			{Crow}}\ and\ \bibinfo {author} {\bibfnamefont {R.}~\bibnamefont {Joynt}},\
	}\href {\doibase 10.1103/PhysRevA.89.042123} {\bibfield  {journal} {\bibinfo
			{journal} {Phys. Rev. A}\ }\textbf {\bibinfo {volume} {89}},\ \bibinfo
		{pages} {042123} (\bibinfo {year} {2014})}\BibitemShut {NoStop}%
	\bibitem [{\citenamefont {Sung}\ \emph {et~al.}(2019)\citenamefont {Sung},
		\citenamefont {Beaudoin}, \citenamefont {Norris}, \citenamefont {Yan},
		\citenamefont {Kim}, \citenamefont {Qiu}, \citenamefont {{von L\"uepke}},
		\citenamefont {Yoder}, \citenamefont {Orlando}, \citenamefont {Viola},
		\citenamefont {Gustavsson},\ and\ \citenamefont {Oliver}}]{Sung_arXiv19}%
	\BibitemOpen
	\bibfield  {author} {\bibinfo {author} {\bibfnamefont {Y.}~\bibnamefont
			{Sung}}, \bibinfo {author} {\bibfnamefont {F.}~\bibnamefont {Beaudoin}},
		\bibinfo {author} {\bibfnamefont {L.~M.}\ \bibnamefont {Norris}}, \bibinfo
		{author} {\bibfnamefont {F.}~\bibnamefont {Yan}}, \bibinfo {author}
		{\bibfnamefont {D.~K.}\ \bibnamefont {Kim}}, \bibinfo {author} {\bibfnamefont
			{J.~Y.}\ \bibnamefont {Qiu}}, \bibinfo {author} {\bibfnamefont
			{U.}~\bibnamefont {{von L\"uepke}}}, \bibinfo {author} {\bibfnamefont
			{J.~L.}\ \bibnamefont {Yoder}}, \bibinfo {author} {\bibfnamefont {T.~P.}\
			\bibnamefont {Orlando}}, \bibinfo {author} {\bibfnamefont {L.}~\bibnamefont
			{Viola}}, \bibinfo {author} {\bibfnamefont {S.}~\bibnamefont {Gustavsson}}, \
		and\ \bibinfo {author} {\bibfnamefont {W.~D.}\ \bibnamefont {Oliver}},\
	}\href {http://arxiv.org/abs/1903.01043} {\bibfield  {journal} {\bibinfo
			{journal} {arXiv:1903.01043}\ } (\bibinfo {year} {2019})}\BibitemShut
	{NoStop}%
	\bibitem [{\citenamefont {K{\"u}bler}\ and\ \citenamefont
		{Zeh}(1973)}]{Kuebler_AP73}%
	\BibitemOpen
	\bibfield  {author} {\bibinfo {author} {\bibfnamefont {O.}~\bibnamefont
			{K{\"u}bler}}\ and\ \bibinfo {author} {\bibfnamefont {H.~D.}\ \bibnamefont
			{Zeh}},\ }\href {\doibase 10.1016/0003-4916(73)90040-7} {\bibfield  {journal}
		{\bibinfo  {journal} {Ann.~Phys.}\ }\textbf {\bibinfo {volume} {76}},\
		\bibinfo {pages} {405} (\bibinfo {year} {1973})}\BibitemShut {NoStop}%
	\bibitem [{\citenamefont {Roszak}\ and\ \citenamefont
		{Cywi{\'n}ski}(2015)}]{Roszak_PRA15}%
	\BibitemOpen
	\bibfield  {author} {\bibinfo {author} {\bibfnamefont {K.}~\bibnamefont
			{Roszak}}\ and\ \bibinfo {author} {\bibfnamefont {{\L}.}~\bibnamefont
			{Cywi{\'n}ski}},\ }\href {\doibase 10.1103/PhysRevA.92.032310} {\bibfield
		{journal} {\bibinfo  {journal} {Phys. Rev. A}\ }\textbf {\bibinfo {volume}
			{92}},\ \bibinfo {pages} {032310} (\bibinfo {year} {2015})}\BibitemShut
	{NoStop}%
	\bibitem [{\citenamefont {Eisert}\ and\ \citenamefont
		{Plenio}(2002)}]{Eisert_PRL02}%
	\BibitemOpen
	\bibfield  {author} {\bibinfo {author} {\bibfnamefont {J.}~\bibnamefont
			{Eisert}}\ and\ \bibinfo {author} {\bibfnamefont {M.~B.}\ \bibnamefont
			{Plenio}},\ }\href {\doibase 10.1103/PhysRevLett.89.137902} {\bibfield
		{journal} {\bibinfo  {journal} {Phys. Rev. Lett.}\ }\textbf {\bibinfo
			{volume} {89}},\ \bibinfo {pages} {137902} (\bibinfo {year}
		{2002})}\BibitemShut {NoStop}%
	\bibitem [{\citenamefont {Hilt}\ and\ \citenamefont {Lutz}(2009)}]{Hilt_PRA09}%
	\BibitemOpen
	\bibfield  {author} {\bibinfo {author} {\bibfnamefont {S.}~\bibnamefont
			{Hilt}}\ and\ \bibinfo {author} {\bibfnamefont {E.}~\bibnamefont {Lutz}},\
	}\href {\doibase 10.1103/PhysRevA.79.010101} {\bibfield  {journal} {\bibinfo
			{journal} {Phys. Rev. A}\ }\textbf {\bibinfo {volume} {79}},\ \bibinfo
		{pages} {010101} (\bibinfo {year} {2009})}\BibitemShut {NoStop}%
	\bibitem [{\citenamefont {Maziero}\ \emph {et~al.}(2010)\citenamefont
		{Maziero}, \citenamefont {Werlang}, \citenamefont {Fanchini}, \citenamefont
		{C\'eleri},\ and\ \citenamefont {Serra}}]{Maziero_PRA10}%
	\BibitemOpen
	\bibfield  {author} {\bibinfo {author} {\bibfnamefont {J.}~\bibnamefont
			{Maziero}}, \bibinfo {author} {\bibfnamefont {T.}~\bibnamefont {Werlang}},
		\bibinfo {author} {\bibfnamefont {F.~F.}\ \bibnamefont {Fanchini}}, \bibinfo
		{author} {\bibfnamefont {L.~C.}\ \bibnamefont {C\'eleri}}, \ and\ \bibinfo
		{author} {\bibfnamefont {R.~M.}\ \bibnamefont {Serra}},\ }\href {\doibase
		10.1103/PhysRevA.81.022116} {\bibfield  {journal} {\bibinfo  {journal} {Phys.
				Rev. A}\ }\textbf {\bibinfo {volume} {81}},\ \bibinfo {pages} {022116}
		(\bibinfo {year} {2010})}\BibitemShut {NoStop}%
	\bibitem [{\citenamefont {Pernice}\ and\ \citenamefont
		{Strunz}(2011)}]{Pernice_PRA11}%
	\BibitemOpen
	\bibfield  {author} {\bibinfo {author} {\bibfnamefont {A.}~\bibnamefont
			{Pernice}}\ and\ \bibinfo {author} {\bibfnamefont {W.~T.}\ \bibnamefont
			{Strunz}},\ }\href {\doibase 10.1103/PhysRevA.84.062121} {\bibfield
		{journal} {\bibinfo  {journal} {Phys. Rev. A}\ }\textbf {\bibinfo {volume}
			{84}},\ \bibinfo {pages} {062121} (\bibinfo {year} {2011})}\BibitemShut
	{NoStop}%
	\bibitem [{\citenamefont {Werner}(1989)}]{Werner_PRA89}%
	\BibitemOpen
	\bibfield  {author} {\bibinfo {author} {\bibfnamefont {R.~F.}\ \bibnamefont
			{Werner}},\ }\href {\doibase 10.1103/PhysRevA.40.4277} {\bibfield  {journal}
		{\bibinfo  {journal} {Phys. Rev. A}\ }\textbf {\bibinfo {volume} {40}},\
		\bibinfo {pages} {4277} (\bibinfo {year} {1989})}\BibitemShut {NoStop}%
	\bibitem [{\citenamefont {Plenio}\ and\ \citenamefont
		{Virmani}(2007)}]{Plenio_QIC07}%
	\BibitemOpen
	\bibfield  {author} {\bibinfo {author} {\bibfnamefont {M.~B.}\ \bibnamefont
			{Plenio}}\ and\ \bibinfo {author} {\bibfnamefont {S.}~\bibnamefont
			{Virmani}},\ }\href@noop {} {\bibfield  {journal} {\bibinfo  {journal}
			{Quant.~Info.~Comput.}\ }\textbf {\bibinfo {volume} {7}},\ \bibinfo {pages}
		{1} (\bibinfo {year} {2007})}\BibitemShut {NoStop}%
	\bibitem [{\citenamefont {Horodecki}\ \emph {et~al.}(2009)\citenamefont
		{Horodecki}, \citenamefont {Horodecki}, \citenamefont {Horodecki},\ and\
		\citenamefont {Horodecki}}]{Horodecki_RMP09}%
	\BibitemOpen
	\bibfield  {author} {\bibinfo {author} {\bibfnamefont {R.}~\bibnamefont
			{Horodecki}}, \bibinfo {author} {\bibfnamefont {P.}~\bibnamefont
			{Horodecki}}, \bibinfo {author} {\bibfnamefont {M.}~\bibnamefont
			{Horodecki}}, \ and\ \bibinfo {author} {\bibfnamefont {K.}~\bibnamefont
			{Horodecki}},\ }\href {\doibase 10.1103/RevModPhys.81.865} {\bibfield
		{journal} {\bibinfo  {journal} {Rev.\ Mod.\ Phys.}\ }\textbf {\bibinfo
			{volume} {81}},\ \bibinfo {pages} {865} (\bibinfo {year} {2009})}\BibitemShut
	{NoStop}%
	\bibitem [{\citenamefont {Braunstein}\ and\ \citenamefont {van
			Loock}(2005)}]{Braunstein_RMP05}%
	\BibitemOpen
	\bibfield  {author} {\bibinfo {author} {\bibfnamefont {S.~L.}\ \bibnamefont
			{Braunstein}}\ and\ \bibinfo {author} {\bibfnamefont {P.}~\bibnamefont {van
				Loock}},\ }\href {\doibase 10.1103/RevModPhys.77.513} {\bibfield  {journal}
		{\bibinfo  {journal} {Rev. Mod. Phys.}\ }\textbf {\bibinfo {volume} {77}},\
		\bibinfo {pages} {513} (\bibinfo {year} {2005})}\BibitemShut {NoStop}%
	\bibitem [{\citenamefont {Aolita}\ \emph {et~al.}(2015)\citenamefont {Aolita},
		\citenamefont {{de Melo}},\ and\ \citenamefont {Davidovich}}]{Aolita_RPP15}%
	\BibitemOpen
	\bibfield  {author} {\bibinfo {author} {\bibfnamefont {L.}~\bibnamefont
			{Aolita}}, \bibinfo {author} {\bibfnamefont {F.}~\bibnamefont {{de Melo}}}, \
		and\ \bibinfo {author} {\bibfnamefont {L.}~\bibnamefont {Davidovich}},\
	}\href {\doibase 10.1088/0034-4885/78/4/042001} {\bibfield  {journal}
		{\bibinfo  {journal} {Rep. Prog. Phys.}\ }\textbf {\bibinfo {volume} {78}},\
		\bibinfo {pages} {042001} (\bibinfo {year} {2015})}\BibitemShut {NoStop}%
	\bibitem [{\citenamefont {Roszak}\ and\ \citenamefont
		{Cywi\ifmmode~\acute{n}\else \'{n}\fi{}ski}(2018)}]{Roszak_PRA18}%
	\BibitemOpen
	\bibfield  {author} {\bibinfo {author} {\bibfnamefont {K.}~\bibnamefont
			{Roszak}}\ and\ \bibinfo {author} {\bibfnamefont {L.}~\bibnamefont
			{Cywi\ifmmode~\acute{n}\else \'{n}\fi{}ski}},\ }\href {\doibase
		10.1103/PhysRevA.97.012306} {\bibfield  {journal} {\bibinfo  {journal} {Phys.
				Rev. A}\ }\textbf {\bibinfo {volume} {97}},\ \bibinfo {pages} {012306}
		(\bibinfo {year} {2018})}\BibitemShut {NoStop}%
	\bibitem [{Note1()}]{Note1}%
	\BibitemOpen
	\bibinfo {note} {Note that this does not necessarily preclude using a
		classical noise picture for calculation of the dephasing, but one should keep
		in mind that this is then only a calculation tool.}\BibitemShut {Stop}%
	\bibitem [{\citenamefont {Wrachtrup}\ and\ \citenamefont
		{Finkler}(2016)}]{Wrachtrup_JMR16}%
	\BibitemOpen
	\bibfield  {author} {\bibinfo {author} {\bibfnamefont {J.}~\bibnamefont
			{Wrachtrup}}\ and\ \bibinfo {author} {\bibfnamefont {A.}~\bibnamefont
			{Finkler}},\ }\href {\doibase 10.1016/j.jmr.2016.06.017} {\bibfield
		{journal} {\bibinfo  {journal} {J.~Magn.~Res.}\ }\textbf {\bibinfo {volume}
			{369}},\ \bibinfo {pages} {225} (\bibinfo {year} {2016})}\BibitemShut
	{NoStop}%
	\bibitem [{\citenamefont {Cywi{\'n}ski}(2011)}]{Cywinski_APPA11}%
	\BibitemOpen
	\bibfield  {author} {\bibinfo {author} {\bibfnamefont {{\L}.}~\bibnamefont
			{Cywi{\'n}ski}},\ }\href@noop {} {\bibfield  {journal} {\bibinfo  {journal}
			{Acta Phys.~Pol.~A}\ }\textbf {\bibinfo {volume} {119}},\ \bibinfo {pages}
		{576} (\bibinfo {year} {2011})}\BibitemShut {NoStop}%
	\bibitem [{\citenamefont {Yang}\ \emph {et~al.}(2017)\citenamefont {Yang},
		\citenamefont {Ma},\ and\ \citenamefont {Liu}}]{Yang_RPP17}%
	\BibitemOpen
	\bibfield  {author} {\bibinfo {author} {\bibfnamefont {W.}~\bibnamefont
			{Yang}}, \bibinfo {author} {\bibfnamefont {W.-L.}\ \bibnamefont {Ma}}, \ and\
		\bibinfo {author} {\bibfnamefont {R.-B.}\ \bibnamefont {Liu}},\ }\href
	{\doibase 10.1088/0034-4885/80/1/016001} {\bibfield  {journal} {\bibinfo
			{journal} {Rep. Prog. Phys.}\ }\textbf {\bibinfo {volume} {80}},\ \bibinfo
		{pages} {016001} (\bibinfo {year} {2017})}\BibitemShut {NoStop}%
	\bibitem [{\citenamefont {Malinowski}\ \emph
		{et~al.}(2017{\natexlab{b}})\citenamefont {Malinowski}, \citenamefont
		{Martins}, \citenamefont {Nissen}, \citenamefont {Barnes}, \citenamefont
		{Cywi\'n{}ski}, \citenamefont {Rudner}, \citenamefont {Fallahi},
		\citenamefont {Gardner}, \citenamefont {Manfra}, \citenamefont {Marcus},\
		and\ \citenamefont {Kuemmeth}}]{Malinowski_NN17}%
	\BibitemOpen
	\bibfield  {author} {\bibinfo {author} {\bibfnamefont {F.~K.}\ \bibnamefont
			{Malinowski}}, \bibinfo {author} {\bibfnamefont {F.}~\bibnamefont {Martins}},
		\bibinfo {author} {\bibfnamefont {P.~D.}\ \bibnamefont {Nissen}}, \bibinfo
		{author} {\bibfnamefont {E.}~\bibnamefont {Barnes}}, \bibinfo {author}
		{\bibfnamefont {L.}~\bibnamefont {Cywi\'n{}ski}}, \bibinfo {author}
		{\bibfnamefont {M.~S.}\ \bibnamefont {Rudner}}, \bibinfo {author}
		{\bibfnamefont {S.}~\bibnamefont {Fallahi}}, \bibinfo {author} {\bibfnamefont
			{G.~C.}\ \bibnamefont {Gardner}}, \bibinfo {author} {\bibfnamefont {M.~J.}\
			\bibnamefont {Manfra}}, \bibinfo {author} {\bibfnamefont {C.~M.}\
			\bibnamefont {Marcus}}, \ and\ \bibinfo {author} {\bibfnamefont
			{F.}~\bibnamefont {Kuemmeth}},\ }\href {\doibase 10.1038/nnano.2016.170}
	{\bibfield  {journal} {\bibinfo  {journal} {Nature Nanotechnology}\ }\textbf
		{\bibinfo {volume} {12}},\ \bibinfo {pages} {16} (\bibinfo {year}
		{2017}{\natexlab{b}})}\BibitemShut {NoStop}%
	\bibitem [{\citenamefont {Zhao}\ \emph {et~al.}(2012)\citenamefont {Zhao},
		\citenamefont {Ho},\ and\ \citenamefont {Liu}}]{Zhao_PRB12}%
	\BibitemOpen
	\bibfield  {author} {\bibinfo {author} {\bibfnamefont {N.}~\bibnamefont
			{Zhao}}, \bibinfo {author} {\bibfnamefont {S.-W.}\ \bibnamefont {Ho}}, \ and\
		\bibinfo {author} {\bibfnamefont {R.-B.}\ \bibnamefont {Liu}},\ }\href
	{\doibase 10.1103/PhysRevB.85.115303} {\bibfield  {journal} {\bibinfo
			{journal} {Phys. Rev. B}\ }\textbf {\bibinfo {volume} {85}},\ \bibinfo
		{pages} {115303} (\bibinfo {year} {2012})}\BibitemShut {NoStop}%
	\bibitem [{\citenamefont {{de Lange}}\ \emph {et~al.}(2010)\citenamefont {{de
				Lange}}, \citenamefont {Wang}, \citenamefont {Rist{\`e}}, \citenamefont
		{Dobrovitski},\ and\ \citenamefont {Hanson}}]{deLange_Science10}%
	\BibitemOpen
	\bibfield  {author} {\bibinfo {author} {\bibfnamefont {G.}~\bibnamefont {{de
					Lange}}}, \bibinfo {author} {\bibfnamefont {Z.~H.}\ \bibnamefont {Wang}},
		\bibinfo {author} {\bibfnamefont {D.}~\bibnamefont {Rist{\`e}}}, \bibinfo
		{author} {\bibfnamefont {V.~V.}\ \bibnamefont {Dobrovitski}}, \ and\ \bibinfo
		{author} {\bibfnamefont {R.}~\bibnamefont {Hanson}},\ }\href {\doibase
		10.1126/science.1192739} {\bibfield  {journal} {\bibinfo  {journal}
			{Science}\ }\textbf {\bibinfo {volume} {330}},\ \bibinfo {pages} {60}
		(\bibinfo {year} {2010})}\BibitemShut {NoStop}%
	\bibitem [{\citenamefont {Witzel}\ \emph {et~al.}(2012)\citenamefont {Witzel},
		\citenamefont {Carroll}, \citenamefont {Cywi{\'n}ski},\ and\ \citenamefont
		{Das~Sarma}}]{Witzel_PRB12}%
	\BibitemOpen
	\bibfield  {author} {\bibinfo {author} {\bibfnamefont {W.~M.}\ \bibnamefont
			{Witzel}}, \bibinfo {author} {\bibfnamefont {M.~S.}\ \bibnamefont {Carroll}},
		\bibinfo {author} {\bibfnamefont {{\L}.}~\bibnamefont {Cywi{\'n}ski}}, \ and\
		\bibinfo {author} {\bibfnamefont {S.}~\bibnamefont {Das~Sarma}},\ }\href
	{\doibase 10.1103/PhysRevB.86.035452} {\bibfield  {journal} {\bibinfo
			{journal} {Phys.\ Rev.\ B}\ }\textbf {\bibinfo {volume} {86}},\ \bibinfo
		{pages} {035452} (\bibinfo {year} {2012})}\BibitemShut {NoStop}%
	\bibitem [{\citenamefont {Witzel}\ \emph {et~al.}(2010)\citenamefont {Witzel},
		\citenamefont {Carroll}, \citenamefont {Morello}, \citenamefont
		{Cywi{\'n}ski},\ and\ \citenamefont {{Das Sarma}}}]{Witzel_PRL10}%
	\BibitemOpen
	\bibfield  {author} {\bibinfo {author} {\bibfnamefont {W.~M.}\ \bibnamefont
			{Witzel}}, \bibinfo {author} {\bibfnamefont {M.~S.}\ \bibnamefont {Carroll}},
		\bibinfo {author} {\bibfnamefont {A.}~\bibnamefont {Morello}}, \bibinfo
		{author} {\bibfnamefont {{\L}.}~\bibnamefont {Cywi{\'n}ski}}, \ and\ \bibinfo
		{author} {\bibfnamefont {S.}~\bibnamefont {{Das Sarma}}},\ }\href {\doibase
		10.1103/PhysRevLett.105.187602} {\bibfield  {journal} {\bibinfo  {journal}
			{Phys.\ Rev.\ Lett.}\ }\textbf {\bibinfo {volume} {105}},\ \bibinfo {pages}
		{187602} (\bibinfo {year} {2010})}\BibitemShut {NoStop}%
	\bibitem [{\citenamefont {Monz}\ \emph {et~al.}(2011)\citenamefont {Monz},
		\citenamefont {Schindler}, \citenamefont {Barreiro}, \citenamefont {Chwalla},
		\citenamefont {Nigg}, \citenamefont {Coish}, \citenamefont {Harlander},
		\citenamefont {H\"ansel}, \citenamefont {Hennrich},\ and\ \citenamefont
		{Blatt}}]{Monz_PRL11}%
	\BibitemOpen
	\bibfield  {author} {\bibinfo {author} {\bibfnamefont {T.}~\bibnamefont
			{Monz}}, \bibinfo {author} {\bibfnamefont {P.}~\bibnamefont {Schindler}},
		\bibinfo {author} {\bibfnamefont {J.~T.}\ \bibnamefont {Barreiro}}, \bibinfo
		{author} {\bibfnamefont {M.}~\bibnamefont {Chwalla}}, \bibinfo {author}
		{\bibfnamefont {D.}~\bibnamefont {Nigg}}, \bibinfo {author} {\bibfnamefont
			{W.~A.}\ \bibnamefont {Coish}}, \bibinfo {author} {\bibfnamefont
			{M.}~\bibnamefont {Harlander}}, \bibinfo {author} {\bibfnamefont
			{W.}~\bibnamefont {H\"ansel}}, \bibinfo {author} {\bibfnamefont
			{M.}~\bibnamefont {Hennrich}}, \ and\ \bibinfo {author} {\bibfnamefont
			{R.}~\bibnamefont {Blatt}},\ }\href {\doibase 10.1103/PhysRevLett.106.130506}
	{\bibfield  {journal} {\bibinfo  {journal} {Phys. Rev. Lett.}\ }\textbf
		{\bibinfo {volume} {106}},\ \bibinfo {pages} {130506} (\bibinfo {year}
		{2011})}\BibitemShut {NoStop}%
	\bibitem [{\citenamefont {Roszak}\ and\ \citenamefont
		{Machnikowski}(2006{\natexlab{b}})}]{Roszak_PLA06}%
	\BibitemOpen
	\bibfield  {author} {\bibinfo {author} {\bibfnamefont {K.}~\bibnamefont
			{Roszak}}\ and\ \bibinfo {author} {\bibfnamefont {P.}~\bibnamefont
			{Machnikowski}},\ }\href {\doibase 10.1016/j.physleta.2005.11.012} {\bibfield
		{journal} {\bibinfo  {journal} {Phys.~Lett.~A}\ }\textbf {\bibinfo {volume}
			{351}},\ \bibinfo {pages} {251} (\bibinfo {year}
		{2006}{\natexlab{b}})}\BibitemShut {NoStop}%
	\bibitem [{\citenamefont {Krzywda}\ and\ \citenamefont
		{Roszak}(2016)}]{Krzywda_SR16}%
	\BibitemOpen
	\bibfield  {author} {\bibinfo {author} {\bibfnamefont {J.}~\bibnamefont
			{Krzywda}}\ and\ \bibinfo {author} {\bibfnamefont {K.}~\bibnamefont
			{Roszak}},\ }\href {\doibase 10.1038/srep23753} {\bibfield  {journal}
		{\bibinfo  {journal} {Sci. Rep.}\ }\textbf {\bibinfo {volume} {6}},\ \bibinfo
		{pages} {23753} (\bibinfo {year} {2016})}\BibitemShut {NoStop}%
	\bibitem [{\citenamefont {Salamon}\ and\ \citenamefont
		{Roszak}(2017)}]{Salamon_PRA17}%
	\BibitemOpen
	\bibfield  {author} {\bibinfo {author} {\bibfnamefont {T.}~\bibnamefont
			{Salamon}}\ and\ \bibinfo {author} {\bibfnamefont {K.}~\bibnamefont
			{Roszak}},\ }\href {\doibase 10.1103/PhysRevA.96.032333} {\bibfield
		{journal} {\bibinfo  {journal} {Phys. Rev. A}\ }\textbf {\bibinfo {volume}
			{96}},\ \bibinfo {pages} {032333} (\bibinfo {year} {2017})}\BibitemShut
	{NoStop}%
	\bibitem [{\citenamefont {Paz-Silva}\ \emph {et~al.}(2017)\citenamefont
		{Paz-Silva}, \citenamefont {Norris},\ and\ \citenamefont
		{Viola}}]{Paz_PRA17}%
	\BibitemOpen
	\bibfield  {author} {\bibinfo {author} {\bibfnamefont {G.~A.}\ \bibnamefont
			{Paz-Silva}}, \bibinfo {author} {\bibfnamefont {L.~M.}\ \bibnamefont
			{Norris}}, \ and\ \bibinfo {author} {\bibfnamefont {L.}~\bibnamefont
			{Viola}},\ }\href {\doibase 10.1103/PhysRevA.95.022121} {\bibfield  {journal}
		{\bibinfo  {journal} {Phys. Rev. A}\ }\textbf {\bibinfo {volume} {95}},\
		\bibinfo {pages} {022121} (\bibinfo {year} {2017})}\BibitemShut {NoStop}%
	\bibitem [{\citenamefont {Makhlin}\ and\ \citenamefont
		{Shnirman}(2004)}]{Makhlin_PRL04}%
	\BibitemOpen
	\bibfield  {author} {\bibinfo {author} {\bibfnamefont {Y.}~\bibnamefont
			{Makhlin}}\ and\ \bibinfo {author} {\bibfnamefont {A.}~\bibnamefont
			{Shnirman}},\ }\href {\doibase 10.1103/PhysRevLett.92.178301} {\bibfield
		{journal} {\bibinfo  {journal} {Phys. Rev. Lett.}\ }\textbf {\bibinfo
			{volume} {92}},\ \bibinfo {pages} {178301} (\bibinfo {year}
		{2004})}\BibitemShut {NoStop}%
	\bibitem [{\citenamefont {Falci}\ \emph {et~al.}(2005)\citenamefont {Falci},
		\citenamefont {D'Arrigo}, \citenamefont {Mastellone},\ and\ \citenamefont
		{Paladino}}]{Falci_PRL05}%
	\BibitemOpen
	\bibfield  {author} {\bibinfo {author} {\bibfnamefont {G.}~\bibnamefont
			{Falci}}, \bibinfo {author} {\bibfnamefont {A.}~\bibnamefont {D'Arrigo}},
		\bibinfo {author} {\bibfnamefont {A.}~\bibnamefont {Mastellone}}, \ and\
		\bibinfo {author} {\bibfnamefont {E.}~\bibnamefont {Paladino}},\ }\href
	{\doibase 10.1103/PhysRevLett.94.167002} {\bibfield  {journal} {\bibinfo
			{journal} {Phys.\ Rev.\ Lett.}\ }\textbf {\bibinfo {volume} {94}},\ \bibinfo
		{pages} {167002} (\bibinfo {year} {2005})}\BibitemShut {NoStop}%
	\bibitem [{\citenamefont {Yao}\ \emph {et~al.}(2006)\citenamefont {Yao},
		\citenamefont {Liu},\ and\ \citenamefont {Sham}}]{Yao_PRB06}%
	\BibitemOpen
	\bibfield  {author} {\bibinfo {author} {\bibfnamefont {W.}~\bibnamefont
			{Yao}}, \bibinfo {author} {\bibfnamefont {R.-B.}\ \bibnamefont {Liu}}, \ and\
		\bibinfo {author} {\bibfnamefont {L.~J.}\ \bibnamefont {Sham}},\ }\href
	{\doibase 10.1103/PhysRevB.74.195301} {\bibfield  {journal} {\bibinfo
			{journal} {Phys.\ Rev.\ B}\ }\textbf {\bibinfo {volume} {74}},\ \bibinfo
		{pages} {195301} (\bibinfo {year} {2006})}\BibitemShut {NoStop}%
	\bibitem [{\citenamefont {Bergli}\ \emph {et~al.}(2006)\citenamefont {Bergli},
		\citenamefont {Galperin},\ and\ \citenamefont {Altshuler}}]{Bergli_PRB06}%
	\BibitemOpen
	\bibfield  {author} {\bibinfo {author} {\bibfnamefont {J.}~\bibnamefont
			{Bergli}}, \bibinfo {author} {\bibfnamefont {Y.~M.}\ \bibnamefont
			{Galperin}}, \ and\ \bibinfo {author} {\bibfnamefont {B.~L.}\ \bibnamefont
			{Altshuler}},\ }\href {\doibase 10.1103/PhysRevB.74.024509} {\bibfield
		{journal} {\bibinfo  {journal} {Phys.\ Rev.\ B}\ }\textbf {\bibinfo {volume}
			{74}},\ \bibinfo {pages} {024509} (\bibinfo {year} {2006})}\BibitemShut
	{NoStop}%
	\bibitem [{\citenamefont {Cywi{\'n}ski}\ \emph {et~al.}(2009)\citenamefont
		{Cywi{\'n}ski}, \citenamefont {Witzel},\ and\ \citenamefont {{Das
				Sarma}}}]{Cywinski_PRB09}%
	\BibitemOpen
	\bibfield  {author} {\bibinfo {author} {\bibfnamefont {{\L}.}~\bibnamefont
			{Cywi{\'n}ski}}, \bibinfo {author} {\bibfnamefont {W.~M.}\ \bibnamefont
			{Witzel}}, \ and\ \bibinfo {author} {\bibfnamefont {S.}~\bibnamefont {{Das
					Sarma}}},\ }\href {\doibase 10.1103/PhysRevB.79.245314} {\bibfield  {journal}
		{\bibinfo  {journal} {Phys.\ Rev.\ B}\ }\textbf {\bibinfo {volume} {79}},\
		\bibinfo {pages} {245314} (\bibinfo {year} {2009})}\BibitemShut {NoStop}%
	\bibitem [{\citenamefont {Ramon}(2012)}]{Ramon_PRB12}%
	\BibitemOpen
	\bibfield  {author} {\bibinfo {author} {\bibfnamefont {G.}~\bibnamefont
			{Ramon}},\ }\href {\doibase 10.1103/PhysRevB.86.125317} {\bibfield  {journal}
		{\bibinfo  {journal} {Phys.\ Rev.\ B}\ }\textbf {\bibinfo {volume} {86}},\
		\bibinfo {pages} {125317} (\bibinfo {year} {2012})}\BibitemShut {NoStop}%
	\bibitem [{\citenamefont {Cywi{\'n}ski}(2014)}]{Cywinski_PRA14}%
	\BibitemOpen
	\bibfield  {author} {\bibinfo {author} {\bibfnamefont {{\L}.}~\bibnamefont
			{Cywi{\'n}ski}},\ }\href {\doibase 10.1103/PhysRevA.90.042307} {\bibfield
		{journal} {\bibinfo  {journal} {Phys. Rev. A}\ }\textbf {\bibinfo {volume}
			{90}},\ \bibinfo {pages} {042307} (\bibinfo {year} {2014})}\BibitemShut
	{NoStop}%
	\bibitem [{\citenamefont {Sza{\'n}kowski}\ \emph {et~al.}(2015)\citenamefont
		{Sza{\'n}kowski}, \citenamefont {Trippenbach}, \citenamefont {Cywi{\'n}ski},\
		and\ \citenamefont {Band}}]{Szankowski_QIP15}%
	\BibitemOpen
	\bibfield  {author} {\bibinfo {author} {\bibfnamefont {P.}~\bibnamefont
			{Sza{\'n}kowski}}, \bibinfo {author} {\bibfnamefont {M.}~\bibnamefont
			{Trippenbach}}, \bibinfo {author} {\bibfnamefont {{\L}.}~\bibnamefont
			{Cywi{\'n}ski}}, \ and\ \bibinfo {author} {\bibfnamefont {Y.~B.}\
			\bibnamefont {Band}},\ }\href {\doibase 10.1007/s11128-015-1044-7} {\bibfield
		{journal} {\bibinfo  {journal} {Quantum Inf.~Process.}\ }\textbf {\bibinfo
			{volume} {14}},\ \bibinfo {pages} {3367} (\bibinfo {year}
		{2015})}\BibitemShut {NoStop}%
	\bibitem [{\citenamefont {DeVience}\ \emph {et~al.}(2015)\citenamefont
		{DeVience}, \citenamefont {Pham}, \citenamefont {Lovchinsky}, \citenamefont
		{Sushkov}, \citenamefont {Bar-Gill}, \citenamefont {Belthangady},
		\citenamefont {Casola}, \citenamefont {Corbett}, \citenamefont {Zhang},
		\citenamefont {Lukin}, \citenamefont {Park}, \citenamefont {Yacoby},\ and\
		\citenamefont {Walsworth}}]{DeVience_NN15}%
	\BibitemOpen
	\bibfield  {author} {\bibinfo {author} {\bibfnamefont {S.~J.}\ \bibnamefont
			{DeVience}}, \bibinfo {author} {\bibfnamefont {L.~M.}\ \bibnamefont {Pham}},
		\bibinfo {author} {\bibfnamefont {I.}~\bibnamefont {Lovchinsky}}, \bibinfo
		{author} {\bibfnamefont {A.~O.}\ \bibnamefont {Sushkov}}, \bibinfo {author}
		{\bibfnamefont {N.}~\bibnamefont {Bar-Gill}}, \bibinfo {author}
		{\bibfnamefont {C.}~\bibnamefont {Belthangady}}, \bibinfo {author}
		{\bibfnamefont {F.}~\bibnamefont {Casola}}, \bibinfo {author} {\bibfnamefont
			{M.}~\bibnamefont {Corbett}}, \bibinfo {author} {\bibfnamefont
			{H.}~\bibnamefont {Zhang}}, \bibinfo {author} {\bibfnamefont
			{M.}~\bibnamefont {Lukin}}, \bibinfo {author} {\bibfnamefont
			{H.}~\bibnamefont {Park}}, \bibinfo {author} {\bibfnamefont {A.}~\bibnamefont
			{Yacoby}}, \ and\ \bibinfo {author} {\bibfnamefont {R.~L.}\ \bibnamefont
			{Walsworth}},\ }\href {\doibase 10.1038/nnano.2014.313} {\bibfield  {journal}
		{\bibinfo  {journal} {Nature Nanotechnology}\ }\textbf {\bibinfo {volume}
			{10}},\ \bibinfo {pages} {129} (\bibinfo {year} {2015})}\BibitemShut
	{NoStop}%
	\bibitem [{\citenamefont {H{\"a}berle}\ \emph {et~al.}(2015)\citenamefont
		{H{\"a}berle}, \citenamefont {Schmid-Lorch}, \citenamefont {Reinhard},\ and\
		\citenamefont {Wrachtrup}}]{Haberle_NN15}%
	\BibitemOpen
	\bibfield  {author} {\bibinfo {author} {\bibfnamefont {T.}~\bibnamefont
			{H{\"a}berle}}, \bibinfo {author} {\bibfnamefont {D.}~\bibnamefont
			{Schmid-Lorch}}, \bibinfo {author} {\bibfnamefont {F.}~\bibnamefont
			{Reinhard}}, \ and\ \bibinfo {author} {\bibfnamefont {J.}~\bibnamefont
			{Wrachtrup}},\ }\href {\doibase 10.1038/nnano.2014.299} {\bibfield  {journal}
		{\bibinfo  {journal} {Nature Nanotechnology}\ }\textbf {\bibinfo {volume}
			{10}},\ \bibinfo {pages} {125} (\bibinfo {year} {2015})}\BibitemShut
	{NoStop}%
	\bibitem [{\citenamefont {Lovchinsky}\ \emph {et~al.}(2016)\citenamefont
		{Lovchinsky}, \citenamefont {Sushkov}, \citenamefont {Urbach}, \citenamefont
		{{de Leon}}, \citenamefont {Choi}, \citenamefont {Greve}, \citenamefont
		{Evans}, \citenamefont {Gertner}, \citenamefont {Bersin}, \citenamefont
		{M{\"u}ller}, \citenamefont {McGuinness}, \citenamefont {Jelezko},
		\citenamefont {Walsworth}, \citenamefont {Park},\ and\ \citenamefont
		{Lukin}}]{Lovchinsky_Science16}%
	\BibitemOpen
	\bibfield  {author} {\bibinfo {author} {\bibfnamefont {I.}~\bibnamefont
			{Lovchinsky}}, \bibinfo {author} {\bibfnamefont {A.~O.}\ \bibnamefont
			{Sushkov}}, \bibinfo {author} {\bibfnamefont {E.}~\bibnamefont {Urbach}},
		\bibinfo {author} {\bibfnamefont {N.~P.}\ \bibnamefont {{de Leon}}}, \bibinfo
		{author} {\bibfnamefont {S.}~\bibnamefont {Choi}}, \bibinfo {author}
		{\bibfnamefont {K.~D.}\ \bibnamefont {Greve}}, \bibinfo {author}
		{\bibfnamefont {R.}~\bibnamefont {Evans}}, \bibinfo {author} {\bibfnamefont
			{R.}~\bibnamefont {Gertner}}, \bibinfo {author} {\bibfnamefont
			{E.}~\bibnamefont {Bersin}}, \bibinfo {author} {\bibfnamefont
			{C.}~\bibnamefont {M{\"u}ller}}, \bibinfo {author} {\bibfnamefont
			{L.}~\bibnamefont {McGuinness}}, \bibinfo {author} {\bibfnamefont
			{F.}~\bibnamefont {Jelezko}}, \bibinfo {author} {\bibfnamefont {R.~L.}\
			\bibnamefont {Walsworth}}, \bibinfo {author} {\bibfnamefont {H.}~\bibnamefont
			{Park}}, \ and\ \bibinfo {author} {\bibfnamefont {M.~D.}\ \bibnamefont
			{Lukin}},\ }\href {\doibase 10.1126/science.aad8022} {\bibfield  {journal}
		{\bibinfo  {journal} {Science}\ }\textbf {\bibinfo {volume} {351}},\ \bibinfo
		{pages} {836} (\bibinfo {year} {2016})}\BibitemShut {NoStop}%
	\bibitem [{\citenamefont {Gali}\ \emph {et~al.}(2008)\citenamefont {Gali},
		\citenamefont {Fyta},\ and\ \citenamefont {Kaxiras}}]{Gali2008}%
	\BibitemOpen
	\bibfield  {author} {\bibinfo {author} {\bibfnamefont {A.}~\bibnamefont
			{Gali}}, \bibinfo {author} {\bibfnamefont {M.}~\bibnamefont {Fyta}}, \ and\
		\bibinfo {author} {\bibfnamefont {E.}~\bibnamefont {Kaxiras}},\ }\href
	{\doibase 10.1103/PhysRevB.77.155206} {\bibfield  {journal} {\bibinfo
			{journal} {Physical Review B}\ }\textbf {\bibinfo {volume} {77}},\ \bibinfo
		{pages} {155206} (\bibinfo {year} {2008})}\BibitemShut {NoStop}%
	\bibitem [{\citenamefont {Yang}\ and\ \citenamefont
		{Liu}(2008)}]{Yang_CCE_PRB08}%
	\BibitemOpen
	\bibfield  {author} {\bibinfo {author} {\bibfnamefont {W.}~\bibnamefont
			{Yang}}\ and\ \bibinfo {author} {\bibfnamefont {R.-B.}\ \bibnamefont {Liu}},\
	}\href {\doibase 10.1103/PhysRevB.78.085315} {\bibfield  {journal} {\bibinfo
			{journal} {Phys.\ Rev.\ B}\ }\textbf {\bibinfo {volume} {78}},\ \bibinfo
		{pages} {085315} (\bibinfo {year} {2008})}\BibitemShut {NoStop}%
	\bibitem [{\citenamefont {Kwiatkowski}\ and\ \citenamefont
		{Cywi\'{n}ski}(2018)}]{Kwiatkowski_PRB18}%
	\BibitemOpen
	\bibfield  {author} {\bibinfo {author} {\bibfnamefont {D.}~\bibnamefont
			{Kwiatkowski}}\ and\ \bibinfo {author} {\bibfnamefont {L.}~\bibnamefont
			{Cywi\'{n}ski}},\ }\href {\doibase 10.1103/PhysRevB.98.155202} {\bibfield
		{journal} {\bibinfo  {journal} {Phys. Rev. B}\ }\textbf {\bibinfo {volume}
			{98}},\ \bibinfo {pages} {155202} (\bibinfo {year} {2018})}\BibitemShut
	{NoStop}%
	\bibitem [{\citenamefont {London}\ \emph {et~al.}(2013)\citenamefont {London},
		\citenamefont {Scheuer}, \citenamefont {Cai}, \citenamefont {Schwarz},
		\citenamefont {Retzker}, \citenamefont {Plenio}, \citenamefont {Katagiri},
		\citenamefont {Teraji}, \citenamefont {Koizumi}, \citenamefont {Isoya},
		\citenamefont {Fischer}, \citenamefont {McGuinness}, \citenamefont
		{Naydenov},\ and\ \citenamefont {Jelezko}}]{London2013}%
	\BibitemOpen
	\bibfield  {author} {\bibinfo {author} {\bibfnamefont {P.}~\bibnamefont
			{London}}, \bibinfo {author} {\bibfnamefont {J.}~\bibnamefont {Scheuer}},
		\bibinfo {author} {\bibfnamefont {J.~M.}\ \bibnamefont {Cai}}, \bibinfo
		{author} {\bibfnamefont {I.}~\bibnamefont {Schwarz}}, \bibinfo {author}
		{\bibfnamefont {A.}~\bibnamefont {Retzker}}, \bibinfo {author} {\bibfnamefont
			{M.~B.}\ \bibnamefont {Plenio}}, \bibinfo {author} {\bibfnamefont
			{M.}~\bibnamefont {Katagiri}}, \bibinfo {author} {\bibfnamefont
			{T.}~\bibnamefont {Teraji}}, \bibinfo {author} {\bibfnamefont
			{S.}~\bibnamefont {Koizumi}}, \bibinfo {author} {\bibfnamefont
			{J.}~\bibnamefont {Isoya}}, \bibinfo {author} {\bibfnamefont
			{R.}~\bibnamefont {Fischer}}, \bibinfo {author} {\bibfnamefont {L.~P.}\
			\bibnamefont {McGuinness}}, \bibinfo {author} {\bibfnamefont
			{B.}~\bibnamefont {Naydenov}}, \ and\ \bibinfo {author} {\bibfnamefont
			{F.}~\bibnamefont {Jelezko}},\ }\href {\doibase
		10.1103/PhysRevLett.111.067601} {\bibfield  {journal} {\bibinfo  {journal}
			{Phys.~Rev.~Lett.}\ }\textbf {\bibinfo {volume} {111}},\ \bibinfo {pages}
		{067601} (\bibinfo {year} {2013})},\ \Eprint {http://arxiv.org/abs/1304.4709}
	{arXiv:1304.4709} \BibitemShut {NoStop}%
	\bibitem [{\citenamefont {Fischer}\ \emph {et~al.}(2013)\citenamefont
		{Fischer}, \citenamefont {Bretschneider}, \citenamefont {London},
		\citenamefont {Budker}, \citenamefont {Gershoni},\ and\ \citenamefont
		{Frydman}}]{Fischer2013a}%
	\BibitemOpen
	\bibfield  {author} {\bibinfo {author} {\bibfnamefont {R.}~\bibnamefont
			{Fischer}}, \bibinfo {author} {\bibfnamefont {C.~O.}\ \bibnamefont
			{Bretschneider}}, \bibinfo {author} {\bibfnamefont {P.}~\bibnamefont
			{London}}, \bibinfo {author} {\bibfnamefont {D.}~\bibnamefont {Budker}},
		\bibinfo {author} {\bibfnamefont {D.}~\bibnamefont {Gershoni}}, \ and\
		\bibinfo {author} {\bibfnamefont {L.}~\bibnamefont {Frydman}},\ }\href
	{\doibase 10.1103/PhysRevLett.111.057601} {\bibfield  {journal} {\bibinfo
			{journal} {Physical Review Letters}\ }\textbf {\bibinfo {volume} {111}},\
		\bibinfo {pages} {057601} (\bibinfo {year} {2013})},\ \Eprint
	{http://arxiv.org/abs/arXiv:1211.5801v1} {arXiv:arXiv:1211.5801v1}
	\BibitemShut {NoStop}%
	\bibitem [{\citenamefont {Pagliero}\ \emph {et~al.}(2018)\citenamefont
		{Pagliero}, \citenamefont {Rao}, \citenamefont {Zangara}, \citenamefont
		{Dhomkar}, \citenamefont {Wong}, \citenamefont {Abril}, \citenamefont
		{Aslam}, \citenamefont {Parker}, \citenamefont {King}, \citenamefont
		{Avalos}, \citenamefont {Ajoy}, \citenamefont {Wrachtrup}, \citenamefont
		{Pines},\ and\ \citenamefont {Meriles}}]{Pagliero2018}%
	\BibitemOpen
	\bibfield  {author} {\bibinfo {author} {\bibfnamefont {D.}~\bibnamefont
			{Pagliero}}, \bibinfo {author} {\bibfnamefont {K.~R.~K.}\ \bibnamefont
			{Rao}}, \bibinfo {author} {\bibfnamefont {P.~R.}\ \bibnamefont {Zangara}},
		\bibinfo {author} {\bibfnamefont {S.}~\bibnamefont {Dhomkar}}, \bibinfo
		{author} {\bibfnamefont {H.~H.}\ \bibnamefont {Wong}}, \bibinfo {author}
		{\bibfnamefont {A.}~\bibnamefont {Abril}}, \bibinfo {author} {\bibfnamefont
			{N.}~\bibnamefont {Aslam}}, \bibinfo {author} {\bibfnamefont
			{A.}~\bibnamefont {Parker}}, \bibinfo {author} {\bibfnamefont
			{J.}~\bibnamefont {King}}, \bibinfo {author} {\bibfnamefont {C.~E.}\
			\bibnamefont {Avalos}}, \bibinfo {author} {\bibfnamefont {A.}~\bibnamefont
			{Ajoy}}, \bibinfo {author} {\bibfnamefont {J.}~\bibnamefont {Wrachtrup}},
		\bibinfo {author} {\bibfnamefont {A.}~\bibnamefont {Pines}}, \ and\ \bibinfo
		{author} {\bibfnamefont {C.~A.}\ \bibnamefont {Meriles}},\ }\href {\doibase
		10.1103/PhysRevB.97.024422} {\bibfield  {journal} {\bibinfo  {journal} {Phys.
				Rev. B}\ }\textbf {\bibinfo {volume} {97}},\ \bibinfo {pages} {024422}
		(\bibinfo {year} {2018})}\BibitemShut {NoStop}%
	\bibitem [{\citenamefont {Wunderlich}\ \emph {et~al.}(2017)\citenamefont
		{Wunderlich}, \citenamefont {Kohlrautz}, \citenamefont {Abel}, \citenamefont
		{Haase},\ and\ \citenamefont {Meijer}}]{Wunderlich2017}%
	\BibitemOpen
	\bibfield  {author} {\bibinfo {author} {\bibfnamefont {R.}~\bibnamefont
			{Wunderlich}}, \bibinfo {author} {\bibfnamefont {J.}~\bibnamefont
			{Kohlrautz}}, \bibinfo {author} {\bibfnamefont {B.}~\bibnamefont {Abel}},
		\bibinfo {author} {\bibfnamefont {J.}~\bibnamefont {Haase}}, \ and\ \bibinfo
		{author} {\bibfnamefont {J.}~\bibnamefont {Meijer}},\ }\href {\doibase
		10.1103/PhysRevB.96.220407} {\bibfield  {journal} {\bibinfo  {journal}
			{Physical Review B}\ }\textbf {\bibinfo {volume} {96}},\ \bibinfo {pages}
		{220407(R)} (\bibinfo {year} {2017})},\ \Eprint
	{http://arxiv.org/abs/1703.09243} {arXiv:1703.09243} \BibitemShut {NoStop}%
	\bibitem [{\citenamefont {{\'{A}}lvarez}\ \emph {et~al.}(2015)\citenamefont
		{{\'{A}}lvarez}, \citenamefont {Bretschneider}, \citenamefont {Fischer},
		\citenamefont {London}, \citenamefont {Kanda}, \citenamefont {Onoda},
		\citenamefont {Isoya}, \citenamefont {Gershoni},\ and\ \citenamefont
		{Frydman}}]{Alvarez2015}%
	\BibitemOpen
	\bibfield  {author} {\bibinfo {author} {\bibfnamefont {G.~A.}\ \bibnamefont
			{{\'{A}}lvarez}}, \bibinfo {author} {\bibfnamefont {C.~O.}\ \bibnamefont
			{Bretschneider}}, \bibinfo {author} {\bibfnamefont {R.}~\bibnamefont
			{Fischer}}, \bibinfo {author} {\bibfnamefont {P.}~\bibnamefont {London}},
		\bibinfo {author} {\bibfnamefont {H.}~\bibnamefont {Kanda}}, \bibinfo
		{author} {\bibfnamefont {S.}~\bibnamefont {Onoda}}, \bibinfo {author}
		{\bibfnamefont {J.}~\bibnamefont {Isoya}}, \bibinfo {author} {\bibfnamefont
			{D.}~\bibnamefont {Gershoni}}, \ and\ \bibinfo {author} {\bibfnamefont
			{L.}~\bibnamefont {Frydman}},\ }\href {\doibase 10.1038/ncomms9456}
	{\bibfield  {journal} {\bibinfo  {journal} {Nature Communications}\ }\textbf
		{\bibinfo {volume} {6}},\ \bibinfo {pages} {8456} (\bibinfo {year} {2015})},\
	\Eprint {http://arxiv.org/abs/1412.8635} {arXiv:1412.8635} \BibitemShut
	{NoStop}%
	\bibitem [{\citenamefont {King}\ \emph {et~al.}(2015)\citenamefont {King},
		\citenamefont {Jeong}, \citenamefont {Vassiliou}, \citenamefont {Shin},
		\citenamefont {Page}, \citenamefont {Avalos}, \citenamefont {Wang},\ and\
		\citenamefont {Pines}}]{King2015}%
	\BibitemOpen
	\bibfield  {author} {\bibinfo {author} {\bibfnamefont {J.~P.}\ \bibnamefont
			{King}}, \bibinfo {author} {\bibfnamefont {K.}~\bibnamefont {Jeong}},
		\bibinfo {author} {\bibfnamefont {C.~C.}\ \bibnamefont {Vassiliou}}, \bibinfo
		{author} {\bibfnamefont {C.~S.}\ \bibnamefont {Shin}}, \bibinfo {author}
		{\bibfnamefont {R.~H.}\ \bibnamefont {Page}}, \bibinfo {author}
		{\bibfnamefont {C.~E.}\ \bibnamefont {Avalos}}, \bibinfo {author}
		{\bibfnamefont {H.-J.}\ \bibnamefont {Wang}}, \ and\ \bibinfo {author}
		{\bibfnamefont {A.}~\bibnamefont {Pines}},\ }\href {\doibase
		10.1038/ncomms9965} {\bibfield  {journal} {\bibinfo  {journal} {Nature
				Communications}\ }\textbf {\bibinfo {volume} {6}},\ \bibinfo {pages} {8965}
		(\bibinfo {year} {2015})},\ \Eprint {http://arxiv.org/abs/1501.2897}
	{arXiv:1501.2897} \BibitemShut {NoStop}%
	\bibitem [{\citenamefont {Scheuer}\ \emph {et~al.}(2017)\citenamefont
		{Scheuer}, \citenamefont {Schwartz}, \citenamefont {M{\"{u}}ller},
		\citenamefont {Chen}, \citenamefont {Dhand}, \citenamefont {Plenio},
		\citenamefont {Naydenov},\ and\ \citenamefont {Jelezko}}]{Scheuer2017}%
	\BibitemOpen
	\bibfield  {author} {\bibinfo {author} {\bibfnamefont {J.}~\bibnamefont
			{Scheuer}}, \bibinfo {author} {\bibfnamefont {I.}~\bibnamefont {Schwartz}},
		\bibinfo {author} {\bibfnamefont {S.}~\bibnamefont {M{\"{u}}ller}}, \bibinfo
		{author} {\bibfnamefont {Q.}~\bibnamefont {Chen}}, \bibinfo {author}
		{\bibfnamefont {I.}~\bibnamefont {Dhand}}, \bibinfo {author} {\bibfnamefont
			{M.~B.}\ \bibnamefont {Plenio}}, \bibinfo {author} {\bibfnamefont
			{B.}~\bibnamefont {Naydenov}}, \ and\ \bibinfo {author} {\bibfnamefont
			{F.}~\bibnamefont {Jelezko}},\ }\href {\doibase 10.1103/PhysRevB.96.174436}
	{\bibfield  {journal} {\bibinfo  {journal} {Physical Review B}\ }\textbf
		{\bibinfo {volume} {96}},\ \bibinfo {pages} {174436} (\bibinfo {year}
		{2017})},\ \Eprint {http://arxiv.org/abs/1706.01315} {arXiv:1706.01315}
	\BibitemShut {NoStop}%
	\bibitem [{\citenamefont {Poggiali}\ \emph {et~al.}(2017)\citenamefont
		{Poggiali}, \citenamefont {Cappellaro},\ and\ \citenamefont
		{Fabbri}}]{Poggiali2017}%
	\BibitemOpen
	\bibfield  {author} {\bibinfo {author} {\bibfnamefont {F.}~\bibnamefont
			{Poggiali}}, \bibinfo {author} {\bibfnamefont {P.}~\bibnamefont
			{Cappellaro}}, \ and\ \bibinfo {author} {\bibfnamefont {N.}~\bibnamefont
			{Fabbri}},\ }\href {\doibase 10.1103/PhysRevB.95.195308} {\bibfield
		{journal} {\bibinfo  {journal} {Physical Review B}\ }\textbf {\bibinfo
			{volume} {95}},\ \bibinfo {pages} {195308} (\bibinfo {year} {2017})},\
	\Eprint {http://arxiv.org/abs/1612.04783} {arXiv:1612.04783} \BibitemShut
	{NoStop}%
	\bibitem [{\citenamefont {Hovav}\ \emph {et~al.}(2018)\citenamefont {Hovav},
		\citenamefont {Naydenov}, \citenamefont {Jelezko},\ and\ \citenamefont
		{Bar-Gill}}]{Hovav2018}%
	\BibitemOpen
	\bibfield  {author} {\bibinfo {author} {\bibfnamefont {Y.}~\bibnamefont
			{Hovav}}, \bibinfo {author} {\bibfnamefont {B.}~\bibnamefont {Naydenov}},
		\bibinfo {author} {\bibfnamefont {F.}~\bibnamefont {Jelezko}}, \ and\
		\bibinfo {author} {\bibfnamefont {N.}~\bibnamefont {Bar-Gill}},\ }\href
	{\doibase 10.1103/PhysRevLett.120.060405} {\bibfield  {journal} {\bibinfo
			{journal} {Phys. Rev. Lett.}\ }\textbf {\bibinfo {volume} {120}},\ \bibinfo
		{pages} {060405} (\bibinfo {year} {2018})}\BibitemShut {NoStop}%
	\bibitem [{\citenamefont {Zhao}\ \emph {et~al.}(2011)\citenamefont {Zhao},
		\citenamefont {Wang},\ and\ \citenamefont {Liu}}]{Zhao_PRL11}%
	\BibitemOpen
	\bibfield  {author} {\bibinfo {author} {\bibfnamefont {N.}~\bibnamefont
			{Zhao}}, \bibinfo {author} {\bibfnamefont {Z.-Y.}\ \bibnamefont {Wang}}, \
		and\ \bibinfo {author} {\bibfnamefont {R.-B.}\ \bibnamefont {Liu}},\ }\href
	{\doibase 10.1103/PhysRevLett.106.217205} {\bibfield  {journal} {\bibinfo
			{journal} {Phys.\ Rev.\ Lett.}\ }\textbf {\bibinfo {volume} {106}},\ \bibinfo
		{pages} {217205} (\bibinfo {year} {2011})}\BibitemShut {NoStop}%
\end{thebibliography}
\end{document}